\documentclass{article}
\usepackage{cite} 
\usepackage{amsmath} 
\usepackage{mathtools} 
\usepackage{bm} 
\usepackage{float}
\usepackage{algorithm}
\usepackage{algorithmic}
\usepackage{fullpage}
\usepackage{arydshln}		
\usepackage{authblk}
\usepackage{booktabs}       %
\usepackage{amsmath,amssymb,amsfonts}
\usepackage{enumerate}
\usepackage{textcomp}
\usepackage[font=small]{caption}
\usepackage{newfloat}
\usepackage{listings}

\usepackage{caption} 
\usepackage{graphicx} 

\usepackage{xspace}
\usepackage{xcolor}
\usepackage{color, colortbl}
\usepackage{multirow}
\usepackage{threeparttable}
\usepackage{arydshln}	
\usepackage{marginnote}
\usepackage{pifont}
\usepackage[colorinlistoftodos,prependcaption,textsize=small]{todonotes}

\let\citeleft=(
\let\citeright=)

\newtagform{brackets}{[}{]}
\usetagform{brackets}

\begin{document}

\pdfinfo{
   /Author (AUTHORS)
   /Title (TITLE)
}

\makeatletter 
\renewcommand\@biblabel[1]{#1.} 
\makeatother

\newcommand{\markup}[1]{#1}

\newtheorem{proposition}{Proposition}
\newtheorem{definition}{Definition}
\newtheorem{theorem}{Theorem}
\newtheorem{lemma}{Lemma}
\newtheorem{corollary}{Corollary}
\newtheorem{assumption}{Assumption}
\newtheorem{claim}{Claim}

\newcommand{\USK}[1]{%
	{\textcolor{red}{\textbf{[Bek: \small #1]}}}%
}%

\def\defn{\,\coloneqq\,}
\def\argmin{\mathop{\mathrm{arg\,min}}} 
\def\argmax{\mathop{\mathrm{arg\,max}}}
\def\lim{\mathop{\mathrm{lim}}} 
\def\min{\mathop{\mathrm{min}}} 
\def\max{\mathop{\mathrm{max}}} 
\def\sup{\mathop{\mathrm{sup}}}
\def\inf{\mathop{\mathrm{inf}}}
\def\prox{\mathrm{prox}}
\def\log{\mathrm{log}}
\def\zer{\mathrm{zer}}
\def\fix{\mathrm{fix}}
\def\d{\, \mathrm{d}} 
\def\DnCNNast{{\text{DnCNN}^\ast}}
\def\uin{u_\text{in}}
\def\usc{u_\text{sc}}
\newcommand{\norm}[1]{\left\lVert#1\right\rVert}

\def\Lmax{{L_{\mathrm{\tiny max}}}}
\def\abm{{\bm{a}}}
\def\bbm{{\bm{b}}}
\def\cbm{{\bm{c}}}
\def\ebm{{\bm{e}}}
\def\hbm{{\bm{h}}}
\def\sbm{{\bm{s}}}
\def\xbm{{\bm{x}}}
\def\gbm{{\bm{g}}}
\def\ybm{{\bm{y}}}
\def\zbm{{\bm{z}}}
\def\rbm{{\bm{r}}}
\def\qbm{{\bm{q}}}
\def\nbm{{\bm{n}}}
\def\ubm{{\bm{u}}}
\def\kbm{{\bm{k}}}
\def\Lbm{{\bm{L}}}
\def\vbm{{\bm{v}}}
\def\wbm{{\bm{w}}}
\def\varphibm{{\bm{\varphi}}}
\def\omegabm{{\bm{\omega}}}
\def\thetabm{{\bm{\theta}}}
\def\phibm{{\bm{\phi}}}
\def\zerobm{\bm{0}}
\def\Abm{{\bm{A}}}
\def\Hbm{{\bm{H}}}
\def\Dbm{{\bm{D}}}
\def\Fbm{{\bm{F}}}
\def\Sbm{{\bm{S}}}
\def\Pbm{{\bm{P}}}
\def\Rbm{{\bm{R}}}
\def\Lbm{{\bm{L}}}
\def\Tbm{{\bm{T}}}
\def\Ibm{{\bm{I}}}
\def\Jbm{{\bm{J}}}
\def\xbf{{\mathbf{x}}}
\def\ybf{{\mathbf{y}}}
\def\rbmp{{\bm{r}^\prime}}

\def\xbmast{{\bm{x}^\ast}}
\def\ybmast{{\bm{y}^\ast}}
\def\zbmast{{\bm{z}^\ast}}

\def\xbmhat{{\widehat{\bm{x}}}}
\def\vbmhat{{\widehat{\bm{v}}}}

\def\xbfhat{{\widehat{\mathbf{x}}}}
\def\xbfhat{{\widehat{\bm{x}}}}
\def\Psfhat{{\widehat{\Psf}}}
\def\Gsfhat{{\widehat{\Gsf}}}
\def\Psfhat{{\widehat{\Psf}}}
\def\nablahat{{\widehat{\nabla}}}

\def\Tsf{{\mathrm{T}}}
\def\Ssf{{\mathrm{S}}}
\def\Dsf{{\mathrm{D}}}
\def\Fsf{{\mathrm{F}}}
\def\Nsf{{\mathrm{N}}}
\def\Gsf{{\mathrm{G}}}
\def\Isf{{\mathrm{I}}}
\def\Psf{{\mathrm{P}}}
\def\Rsf{{\mathrm{R}}}
\def\Usf{{\mathrm{U}}}
\def\Hsf{{\mathsf{H}}}
\def\isf{{\mathrm{i}}}
\def\Asf{{\mathrm{A}}}

\def\C{\mathbb{C}}
\def\R{\mathbb{R}}
\def\E{\mathbb{E}}
\def\N{\mathbb{N}}
\def\Z{\mathbb{Z}}

\def\Rcal{{\mathcal{R}}}
\def\Ccal{{\mathcal{C}}}
\def\Ncal{{\mathcal{N}}}
\def\Gcal{{\mathcal{G}}}
\def\Bcal{{\mathcal{B}}}
\def\Acal{{\mathcal{A}}}
\def\Lcal{{\mathcal{L}}}
\def\Pcal{{\mathcal{P}}}

\def\phibm{{\bm{\phi}}}
\def\phibmhat{{\bm{\hat{\phi}}}}
\def\zbmhat{{\bm{\hat{z}}}}
\def\mbm{{\bm{m}}}

\title{\vspace{-2cm} SPICER: Self-Supervised Learning for MRI with Automatic Coil Sensitivity Estimation and Reconstruction}

\author[1,*]{Yuyang~Hu}
\author[2,*]{Weijie~Gan}
\author[3]{Chunwei~Ying}
\author[4]{Tongyao Wang}
\author[3]{Cihat Eldeniz}
\author[1]{Jiaming Liu}
\author[5]{Yasheng Chen}
\author[1,3,4,5]{Hongyu An}
\author[1,2]{Ulugbek~S.~Kamilov}

\affil[1]{\small Department of Electrical and Systems Engineering, Washington University in St.~Louis, St.~Louis, MO 63130, USA}
\affil[2]{\small Department of Computer Science and Engineering, Washington University in St.~Louis, St.~Louis, MO 63130, USA}
\affil[3]{\small Mallinckrodt Institute of Radiology, Washington University in St.~Louis, St.~Louis, MO 63110, USA}
\affil[4]{\small Department of Biomedical Engineering, Washington University in St.~Louis, St.~Louis, MO 63130, USA}
\affil[5]{\small Department of Neurology, Washington University in St.~Louis, St.~Louis, MO 63130, USA}
\affil[*]{\small These authors have contributed equally to the work}
\maketitle

\vfill
\noindent
\textit{Running head:} Self-Supervised Learning for MRI with Automatic Coil Sensitivity Estimation and Reconstruction

\noindent
\textit{Address correspondence to:} \\
Ulugbek~S.~Kamilov, PhD\\ 
 One Brookings Drive\\
 MSC 1045-213-1010J\\
 St.~Louis, MO 63130, USA.\\ 
 Email: kamilov@wustl.com

\noindent
This work was supported in part by the NSF CAREER award under CCF-2043134, NIH R01 EB032713, RF1NS116565, R21NS127425 and NIH R01HL129241.

\noindent
Approximate word count: 250 (Abstract)  3,700 (body)\\

\noindent
Submitted to \textit{Magnetic Resonance in Medicine} as a Research Article.

\clearpage

\section*{Abstract}

\noindent
\textbf{Purpose}: \markup{To introduce a novel deep model-based architecture (DMBA), {\bf SPICER},  that uses pairs of noisy and undersampled k-space measurements of the same object to jointly train a model for MRI reconstruction and automatic coil sensitivity estimation.}

\noindent
\textbf{Methods}: SPICER consists of two modules to simultaneously reconstructs accurate MR images and estimates high-quality coil sensitivity maps (CSMs). The first module, {\bf CSM estimation module}, uses a convolutional neural network (CNN) to estimate CSMs from the raw measurements. The second module, {\bf DMBA-based MRI reconstruction module}, forms reconstructed images from the input measurements and the estimated CSMs using both the physical measurement model and learned CNN prior. With the benefit of our self-supervised learning strategy,
SPICER can be efficiently trained without any fully-sampled reference data.

\noindent
\textbf{Results}: We validate SPICER on both open-access datasets and experimentally collected data, showing that it can achieve state-of-the-art performance in highly accelerated data acquisition settings (up to $10\times$). Our results also highlight the importance of different modules of SPICER---including the DMBA, the CSM estimation, and the SPICER training loss---on the final performance of the method. Moreover, SPICER can estimate better CSMs than pre-estimation methods especially when the ACS data is limited.

\noindent
\textbf{Conclusion}: Despite being trained on noisy undersampled data, SPICER can reconstruct high-quality
images and CSMs in highly undersampled settings, which outperforms other self-supervised learning methods and matches the performance of the well-known E2E-VarNet trained on fully-sampled groundtruth data.
  
\noindent
\textbf{Keywords}: Parallel MRI, Image Reconstruction, Inverse Problems, Deep Learning, Coil Sensitivity Estimation.

\clearpage

\section{Introduction}
\label{sec:introduction}


Magnetic resonance imaging (MRI) is a medical imaging technology known to suffer from slow data acquisition. \emph{Parallel MRI (PMRI)} is a widely-used acceleration strategy that relies on the spatial encoding provided by multiple receiver coils to reduce the amount of data to acquire~\cite{griswoldGeneralized2002, sodicksonSimultaneous1997, pruessmannSENSE1999, ueckerESPIRiTan2014}. The multi-coil under-sampled data can be reconstructed by fitting the missing k-space lines~\cite{griswoldGeneralized2002, sodicksonSimultaneous1997} or in the image space using coil sensitivity maps (CSMs)~\cite{pruessmannSENSE1999,ueckerESPIRiTan2014}. \emph{Compressed sensing (CS)} is a complementary technique used to further accelerate data collection by using prior knowledge on the unknown image (sparsity, low-rankness)~\cite{lustigSparse2007, yeCompressed2019}.  
 
 Deep learning (DL) has recently emerged as a promising paradigm for image reconstruction in CS-PMRI~\cite{ongieDeep2020, knollDeeplearning2020, montalt2021machine}. 
Traditional DL methods train \emph{convolutional neural networks (CNNs)} to map acquired measurements to the desired images~\cite{Jin.etal2017, Zhu.etal2018}.
Recent work has shown that \emph{deep model-based architectures (DMBAs)} can perform better than generic CNNs by accounting for the measurement model of the parallel imaging system~\cite{ hammernikSystematic2021, liangDeep2020, duanVSNet2019, aggarwalMoDL2019, hammernikLearning2018,yaman2020self}.
Most of these methods require pre-calibrated CSMs as an important element in their model.
However, CSM pre-estimation strategies rely on sufficient \emph{auto-calibration signal (ACS)} lines, which limits the acceleration rates for data acquisition. To address this limitation, recent work has proposed to jointly estimate high-quality images and CSMs in an end-to-end manner~\cite{arvinteDeep2021, junJoint2021, sriramEndtoend2020}. However, these methods still require fully-sampled groundtruth images as training targets, which limits their applicability to settings where groundtruth is difficult to obtain or unavailable. On the other hand, there has also been a broad interest in developing self-supervised DL methods that rely exclusively on the information available in the undersampled measurements~\cite{yaman2020selfsupervised, yaman2020self, akcakayaUnsupervised2021, lehtinenNoise2Noise2018, eldenizPhase2Phase2021, ganDeep2021, liuRARE2020}. 

Despite the rich literature on DMBAs and self-supervised DL, the existing work in the area has not investigated joint image reconstruction and coil sensitivity estimation directly from noisy and undersampled data. We bridge this gap by presenting \emph{Self-Supervised Learning for MRI with Automatic Coil Sensitivity Estimation (SPICER)} as a new self-supervised learning framework for parallel MRI that is equipped with an automatic CSM estimator. SPICER is a synergistic combination of a powerful model-based architecture and a flexible self-supervised training scheme. The SPICER architecture consists of two branches: (a) a CNN for estimating CSMs from possibly limited ACS data, while ensuring physically realistic predictions; (b) a DMBA that uses the estimated CSMs for high-quality image reconstruction. The SPICER training is performed using undersampled and noisy measurements without any fully-sampled groundtruth. For training, SPICER necessitates at least one pair of undersampled and noisy measurements from each slice. We extensively validated SPICER on \emph{in-vivo} MRI data for several acceleration factors. Our results show that SPICER can achieve state-of-the-art performance on PMRI at high acceleration rates (up to $10\times$). Moreover, SPICER can estimate better CSMs than pre-estimation methods especially when the ACS data is limited.

This paper extends the preliminary work presented in the workshop paper~\cite{gan2021ss}. Compared to the method in~\cite{gan2021ss}, SPICER uses a different forward model, model-based deep learning architecture, and training loss function. This paper also provides an expanded discussion on related work, additional technical details, as well as completely new numerical results using real in-vivo MRI data acquired using a 32-channel coil.

\section{Theory}
\label{sec:Theory}
\subsection{Problem Formulation}

Consider the following CS-PMRI measurement model
\begin{equation}
\label{equ:inverse_problem}
   \ybm = \Abm\xbm + \ebm\ , 
\end{equation}
where $\xbm\in\C^{n}$ is an unknown image, $\ybm=(\ybm_1,\cdots,\ybm_{n_c})$ are the multi-coil measurements from $n_c \geq 1$ coils, $\ebm=( \ebm_1,\cdots,\ebm_{n_c})$ is the noise vector, and $\Abm = ( \Abm_1,\cdots,\Abm_{n_c})$ is the measurement operator (or forward operator). The measurement model for each coil can be represented as 
\begin{equation}
\label{equ:pmri}
\ybm_k = \underbrace{\Pbm\Fbm\Sbm_k}_{\Abm_k}\xbm + \ebm_k, \quad k=1,2, \dots ,n_c\ ,
\end{equation}
where $\Sbm_k\in\C^{n\times n}$ is the CSM of the $k$th coil, $\Fbm\in\C^{n\times n}$ is the Fourier transform operator, $\Pbm\in\C^{n\times n}$ is the k-space sampling operator, and $\ebm\in\C^{n}$ is the noise vector. Note that $\Sbm=( \Sbm_1,\cdots,\Sbm_{n_c})$ varies for each scan, since it depends on the relative location of the coils with the object being imaged. When $\Sbm$ are known, image reconstruction can be formulated as regularized optimization
\begin{equation}
	\label{equ:optimization}
    \xbmhat = \argmin_\xbm f(\xbm) \quad\text{with}\quad f(\xbm) = g(\xbm) + h(\xbm)\ ,
\end{equation}
where $g$ is the data fidelity term that quantifies consistency with the observed data $\ybm$ and $h$ is a regularizer that infuses prior knowledge on $\xbm$. Examples of $g$ and $h$ used in CS-PMRI are the least-squares and \emph{total variation (TV)} functions~\cite{block2007undersampled} 
\begin{equation}
	\label{equ:optimization_tv}
	g(x)= \frac{1}{2}\norm{\Abm\xbm-\ybm}_2^2 \quad \mathrm{and}  \quad h(x)=\tau\norm{\Dbm\xbm}_1, 
\end{equation}
where $\Dbm$ denotes the image gradient and $\tau > 0$ is a regularization parameter.

DL has recently gained popularity in MRI image reconstruction due to its excellent empirical performance~\cite{ongieDeep2020, knollDeeplearning2020, wangDeep2020}. Traditional DL methods are based on training CNNs (such as U-Net~\cite{ronnebergerUnet2015}) to map the corrupted images~\cite{leeDeep2017, wangAccelerating2016} or the under-sampled measurements~\cite{zhuImage2018, hanKSpace2020} to their desired fully sampled ground-truth versions. There is also growing interest in DMBAs that can combine physical measurement models and learned CNN priors. Well known examples of DMBAs are \emph{plug-and-play priors (PnP)}~\cite{Venkatakrishnan.etal2013, Kamilov.etal2022}, \emph{regularized by denoising (RED)}~\cite{Romano.etal2017}, and \emph{deep unfolding (DU)}~\cite{Schlemper.etal2018, Hammernik.etal2018, Aggarwal.etal2019}. In particular, DU has gained considerable recognition due to its ability to achieve the state-of-the-art performance, while providing robustness to changes in data acquisition. DU architectures are typically obtained by unfolding iterations of an image reconstruction algorithm as layers, representing the regularizer within image reconstruction as a CNN, and training the resulting network end-to-end. Different DU architectures can be obtained by using various optimization/reconstruction algorithms. In this paper, we will rely on a DU variant of the RED model as the basis of our image reconstruction method~\cite{liuSGDNet2021}.  

\subsection{Reconstruction using Pre-Calibrated CSMs}

There are two widely-used image formation approaches in CS-PMRI (see recent review~\cite{knollDeeplearning2020}): (a) reconstruction in the k-space domain and (b) reconstruction in the image domain. GRAPPA~\cite{sodicksonSimultaneous1997} is a well-known example of (a) that fills in unacquired k-space values by linearly interpolating acquired neighboring k-space samples. Recent work~\cite{akccakaya2019scan} extends GRAPPA by using a CNN to learn a non-linear interpolator in k-space. SENSE~\cite{pruessmannSENSE1999} and ESPIRiT~\cite{ueckerESPIRiTan2014} are two well-known examples of (b) that first pre-calibrate CSMs and then use it to solve the inverse problem~\eqref{equ:inverse_problem}. Our work in this paper adopts strategy (b), which will be the focus of the subsequent discussion.

Pre-estimated CSMs can either be obtained by doing a separate calibration scan~\cite{yingJoint2007} or estimated directly from the ACS region of the undersampled measurements. The drawback of the former approach is that it extends the total scan time. ESPIRiT~\cite{ueckerESPIRiTan2014} is based on the latter approach. There are several issues and challenges with the pre-estimated approaches~\cite{ueckerImage2008, yingJoint2007}. One issue is that the inconsistencies between the calibration scan and the accelerated scan can result in imaging artifacts. Another issue is that estimating CSMs from a small number of ACS lines may not be sufficiently accurate. DeepSENSE~\cite{peng2022deepsense}, a recent supervised DL method, uses a CNN to learn a mapping from the ACS data to CSM references, obtained by dividing the fully-sampled individual coil images by the sum-of-squares (SoS) reconstruction from the fully-sampled measurements. While DeepSENSE improves over methods based on pre-estimating CSMs, especially when the ACS data is limited, DeepSENSE still requires fully-sampled data to generate training CSMs.
\subsection{Joint Reconstruction and CSM estimation}

Traditionally, optimization-based methods for joint image reconstruction and CSM estimation treat $\Sbm$ as another unknown variable in \eqref{equ:optimization} and alternate between updating the image and updating the coil sensitivities~\cite{ueckerImage2008, yingJoint2007}. Deep unfolding has recently been adopted to perform joint estimation of image and CSMs without any pre-calibration procedure~\cite{arvinteDeep2021, junJoint2021, sriramEndtoend2020, yiasemis2022recurrent, blumenthal2022nlinv, tang2023jsense, meng2019prior}. The concept behind these methods is to model CSM estimated as a trainable DNN module that can be optimized simultaneously with other learnable parameters in the deep network.
The inputs to the CSM estimated modules could be the original undersampled measurements~\cite{sriramEndtoend2020} or the intermediate results available at different layers of the deep unfolded networks~\cite{arvinteDeep2021, junJoint2021}. However, these joint learning methods rely on fully-sampled groundtruth images as training targets, which limits their applicability when groundtruth data is not available. Our work contributes to this area by investigating a self-supervised learning approach for joint image reconstruction and CSM calibration that requires no fully-sampled groundtruth data.

\subsection{Self-supervised Image Reconstruction}
There is a growing interest in DL-based image reconstruction to reduce the dependence of training on high-quality groundtruth data (see recent reviews~\cite{zeng2021review, Akcakaya.etal2021,Tachella.etal2022a}). Some well-known strategies include \emph{Noise2Noise (N2N)}~\cite{Lehtinen.etal2018}, \emph{Noise2Void (N2V)}~\cite{Krull.etal2019}, \emph{deep image prior (DIP)}~\cite{Ulyanov.etal2018}, \emph{Compressive Sensing using Generative Models (CSGM)}~\cite{bora2018ambientgan}, and equivariant imaging~\cite{chen2021equivariant}. This work is most related to N2N, where a DNN $\mathrm{R}_\thetabm$ is trained on a set of noisy images $\{\xbmhat_{i,j} = \xbm_{i}+ \ebm_{i,j}\}$ with $j$ indexing different realizations of the same underlying image $i$.The N2N learning is formulated as follows
\begin{equation}
	\label{equ:n2n}
	\widehat{\thetabm} = \argmin_\thetabm \sum_i\sum_{j\neq j'}\norm{\mathsf{f}_\thetabm(\ybm_{i,j}) - \ybm_{i,j'}}^2_2\ ,
\end{equation}
where $\mathsf{f}_\thetabm$ denotes a CNN with $\thetabm$ being trainable parameters. N2N and its extensions have been investigated in several papers on PMRI reconstruction~\cite{eldenizPhase2Phase2021, ganDeep2021, liuRARE2020}. For example, DURED-Net~\cite{huang2022unsupervised} proposes an unsupervised learning method for MRI image reconstruction combining an unsupervised denoising network and a PnP method. SSDU and NLINV-Net are related self-supervised deep learning methods for MRI reconstruction that use a similar approach of dividing each MRI acquisition into two subsets for training. However, they have distinct strategies for coil sensitivity map (CSM) estimation. Specifically, SSDU relies on pre-calibrated CSMs, while NLINV-Net adopts a model-based approach for estimating CSMs. Our SPICER method is different from NLINV-Net in that it trains a CNN module to estimate CSMs within the proposed DMBA architecture, thus distinguishing it from the NLINV-Net's pure model-based approach.

While the concept of N2N enables the training of the DNN for PMRI without any fully sampled data, to the best of our knowledge, the prior work is based on using pre-estimated CSMs. Our SPICER method does not require pre-scan calibration, instead using the N2N framework for joint reconstruction and CSM estimation without any groundtruth. In addition, our CSMs are estimated using a learning based strategy.

\begin{figure}[H]
\centering \includegraphics[width=.975\textwidth]{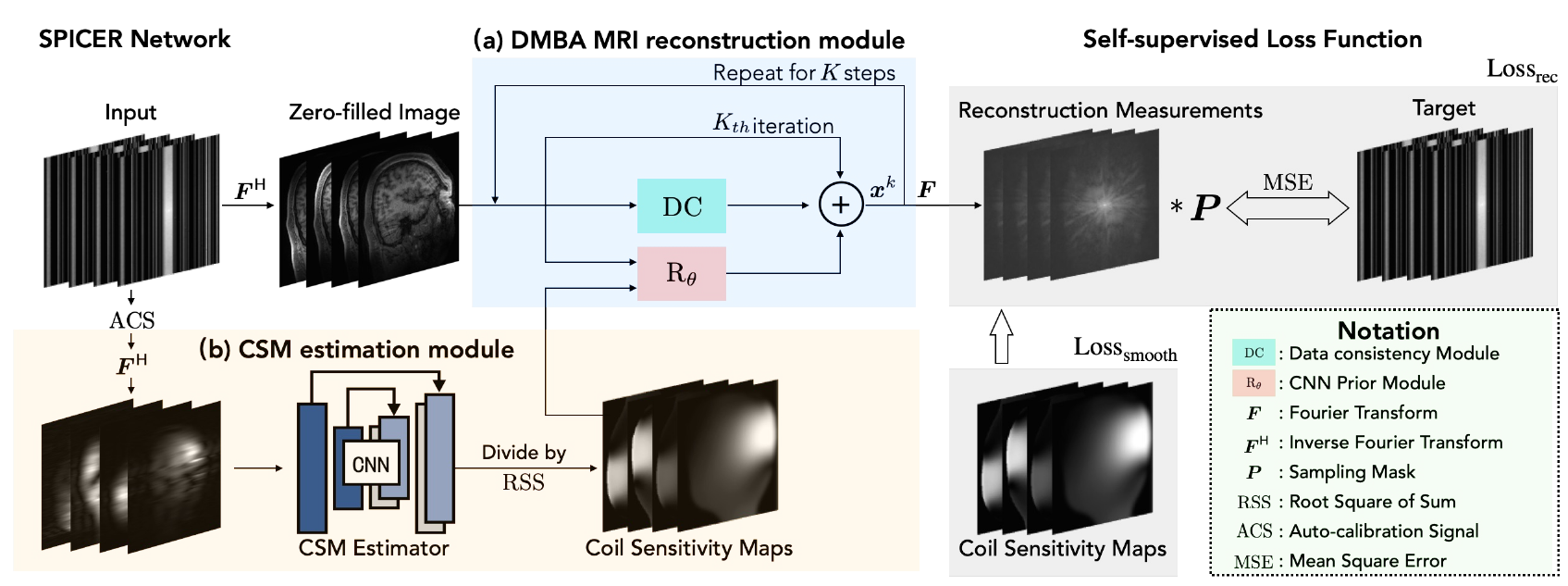} 
	\caption{The SPICER method consists of a DMBA-based MRI reconstruction module and a coil sensitivity estimation module that map multicoil undersampled measurements to a single high-quality image and a set of coil sensitivity maps, respectively. The reconstruction module, described in~\eqref{equ:unfolded}, is composed of two key components, a data consistency module and a CNN prior. The network is trained directly on raw k-space measurements where the input and the target measurement correspond to a pair of undersampled measurements from the same object.}
	\label{fig:method}
\end{figure}

\section{Methods}
\label{sec:methods}

\subsection{SPICER Model}
Our SPICER method takes multicoil undersampled measurements as its input and produces the reconstructed images and CSMs at its output.
As illustrated in Fig.~\ref{fig:method}, SPICER consists of two modules: \emph{(a)} a {\bf CSM estimation module} that uses information extracted from the raw measurements, and \emph{(b)} a {\bf DMBA-based MRI reconstruction module} that forms reconstructed images from the input measurements and the estimated CSMs. Our training procedure uses a pair of multicoil undersampled measurement $\{({\ybm}_i, {\ybm}_i')\}_i$ which are acquired from the same object
\begin{equation}
	\label{equ:pro-fwd}
	\begin{aligned}
		{\ybm}_i = {\Abm}_i \xbm_i + {\ebm}_i\quad \text{and}\quad {\ybm_i'} = {\Abm_i'} \xbm_i + {\ebm_i'}\ ,
	\end{aligned}
\end{equation}
where $({\Abm}_i,  {\Abm}_i' )$ and $({\ebm}_i,  {\ebm}_i' )$ denote distinct forward operators and noise vectors, respectively.
The measurements ${\ybm}_i$ and ${\ybm_i'}$ can correspond to two subsets extracted from a single acquisition~\cite{yaman2020self} or two separate MRI acquisitions. Note that our training procedure does not require any groundtruth images or known CSMs.

\subsection{Coil Sensitivity Estimation Module}

Let $\hat{\ybm}$ be an input measurement and ${\Pbm}$ the corresponding sampling matrix. The coil sensitivity estimation module forms CSMs from multicoil measurements \emph{without} a prescan calibration by performing three steps:
\emph{(a)} the ACS regions of the undersampled measurements are extracted, which are represented as ${\ybm}_{\mathrm{\mathrm{ACS}}}$;
\emph{(b)} ${\ybm}_{\mathrm{\mathrm{ACS}}}$ is mapped back to the image domain by applying the zero-filled inverse Fourier transform ${\bm p}^0 = \Fbm^{-1}({\ybm}_{\mathrm{\mathrm{ACS}}})$;
\emph{(c)} ${\bm p}^0$ is fed into a CNN $\mathrm{P}_\varphibm$ with trainable parameters $\varphibm\in\R^q$ to obtain estimated CSMs: ${\hat{\Sbm}}=\mathrm{P}_\varphibm({\bm p}^0)$. \emph{(d)} Finally, the estimated sensitivity maps ${\hat{\Sbm}}$ are normalized by dividing their root-square-of-sum (RSS) to ensure that ${\hat{\Sbm}}^{-1}{\hat{\Sbm}} = \Ibm$.


\subsection{Image Reconstruction Module}
The image reconstruction module of SPICER incorporates a DMBA, which is built upon RED~\cite{liuSGDNet2021}. Fig. 1 visually represents the DMBA MRI reconstruction module, highlighting two essential steps: the Data Consistency module and the CNN Prior module. The image reconstruction process is performed iteratively by integrating information from CNN $\mathrm{R}_\thetabm$ with learnable parameters $\thetabm$ and imposing consistency between the predicted and the raw k-space measurements via $\nabla g $ in (\ref{equ:data_consistency}).
Let $\hat{\cbm}^0=\Fbm^{-1}\hat{\ybm}$
 represent the initial image, and $K \ge 1$ be the total number of DU layers, The steps of SPICER are given by
\begin{equation}
	\label{equ:unfolded}
	{\cbm}^{k+1} = {\cbm}^{k} - \gamma^{k} (\underbrace{\nabla g({\cbm}^k, \ybm)}_\text{Data Consistency} + \tau^k \underbrace{{\Sbm}\mathrm{R}^k_\thetabm({\Sbm}^{-1}{\cbm}^k)}_\text{Regularization})\ ,
\end{equation} 
where $\tau^k$ and $\gamma^{k}$ are learnable parameters, ${\cbm}^{k}$ are the intermediate \emph{multicoil} images in the $k$th step, and 
\begin{equation}
	\label{equ:data_consistency}
	\nabla g({\cbm}^k, \ybm) = \Fbm^{-1}\Pbm^{-1}({\Pbm}\Fbm{\cbm}^k - \ybm)\ .
\end{equation}
The CNN $\mathrm{R}_\thetabm^k$ in~\eqref{equ:unfolded} takes a single image as an input, requiring ${\Sbm}^{-1}$ and ${\Sbm}$ for fusing multiple images into a single image and expanding a single image into multiple images, respectively. The final reconstructed image ${\xbm}$ can be obtained from the output of the last step as ${\xbm}={\Sbm}^{-1}{\cbm}^K$.
\markup{It is worth noting that unlike existing DU methods that rely on pre-calibrated ${\Sbm}$~\cite{duanVSNet2019, aggarwalMoDL2019, hammernikLearning2018, yaman2020self}, SPICER jointly trains a ${\Sbm}$ calibration network $\mathrm{P}_\varphibm$ simultaneously with the image reconstruction network. During inference,  the pre-trained $\mathrm{P}_\varphibm$ network predicts ${\Sbm}$ as a preliminary step, which is then utilized by the image reconstruction network.}

\subsection{Self-supervised Training Procedure}
We use standard stochastic gradient method to jointly optimize $\{\thetabm_k\}_{k = 1}^K$ and $\varphibm$ by minimizing the loss function
\begin{equation}
	\label{equ:loss}
	\mathrm{Loss} = \mathrm{Loss}_\mathrm{rec} + \lambda\cdot \mathrm{Loss}_\mathrm{smooth}\ ,
\end{equation}
where $\lambda$ is a regularization parameter.

$\mathrm{Loss}_\mathrm{rec}$ seeks to map each ${\ybm}_i$ and the corresponding ${\ybm_i'}$ to each other.
The key idea here is to map the reconstructed images back to the k-space domain by applying the forward operator of the training target.
For example, one can map ${\xbm}_i$ back to the k-space domain by applying the forward operator ${\Abm'}_i$ then penalize the discrepancy between the resulting measurements ${\Abm'}_i{\xbm}_i$ and raw measurements ${\ybm'}_i$. Note that, forward operator ${\Abm'}_i$ uses the same sampling mask $\Pbm'$ employed in the acquisition of the undersampled raw measurement  ${\ybm'}_i$.
Here, the CSMs ${\Sbm}_i$ in ${\Abm}_i$ are estimated by the {coil sensitivity estimation module} after feeding ${\ybm}_i$ as the input.
The formulation of $\mathrm{Loss}_\mathrm{rec}$ is
\begin{equation}
	\label{equ:loss-rec}
	\mathrm{Loss}_\mathrm{rec} = \frac{1}{N} \sum_{i=1}^N\ \Lcal_\mathrm{rec}(\Abm'_i{\xbm}_i\ ,\ \ybm'_i) + \Lcal_\mathrm{rec}({\Abm}_i{\xbm'}_i\ ,\ {\ybm}_i)\ ,
\end{equation}
where $N$ is number of samples, ${\xbm_i'}$ is the reconstructed image when ${\ybm_i'}$ is the input measurement, and $\Lcal_\mathrm{rec}$ denotes the $\ell_2$-norm.
During minimization, $\mathrm{Loss}_\mathrm{rec}$ enforces the accuracy between the predicted and the raw measurements, but it can also generate non-smooth CSMs that are not physically realistic and cause overfitting.
Therefore, we include $\mathrm{Loss}_\mathrm{smooth}$, a smoothness regularization for CSMs, to impose smoothness within the field of view (FOV) region of estimated CSMs.
\begin{equation}
	\label{equ:SmoothingPrior}
	\mathrm{Loss}_\mathrm{smooth} = \frac{1}{N} \sum_{i=1}^N \norm{\Dbm{(\Sbm)}^{\mathrm{FOV}}_i}_2^2,
\end{equation}
where the $\Dbm$ is the gradient of the CSMs ${\Sbm}_i$. The combination of  $\mathrm{Loss}_\mathrm{rec}$ and  $ \mathrm{Loss}_\mathrm{smooth}$ can enable the optimization on the measurement domain and the stability of CSMs, thus it is named as a smoothness-enhanced measurement domain loss function.

\subsection{ Implementation}
In our experiment, all CNNs in SPICER are based on U-Net~\cite{ronnebergerUnet2015} and implemented using PyTorch~\cite{NEURIPS2019_bdbca288}. To process complex-valued data, we reshape all the complex-valued images by splitting their real and
imaginary parts and concatenating them into the channel (= 2) dimensions. The total number of unrolling layers $K$ in \eqref{equ:unfolded} is 8 and the regularization parameter $\lambda$ in \eqref{equ:loss} is 0.01. The initial values we used in \eqref{equ:unfolded} is $\gamma^0 = 1$ and $\tau^0 = 0.1$. We used Adam as the optimizer with a learning rate 0.001 for the initial 30 epochs and 0.0001 for the rest. We performed all our experiments on a machine equipped with an Intel Xeon Gold 6130 Processor and an NVIDIA GeForce RTX 3090 GPU. \markup{The implementation details can be found in: \text{https://github.com/wustl-cig/SPICER.}}

\begin{figure}[H] 
\centering 
\includegraphics[width=.8975\textwidth]{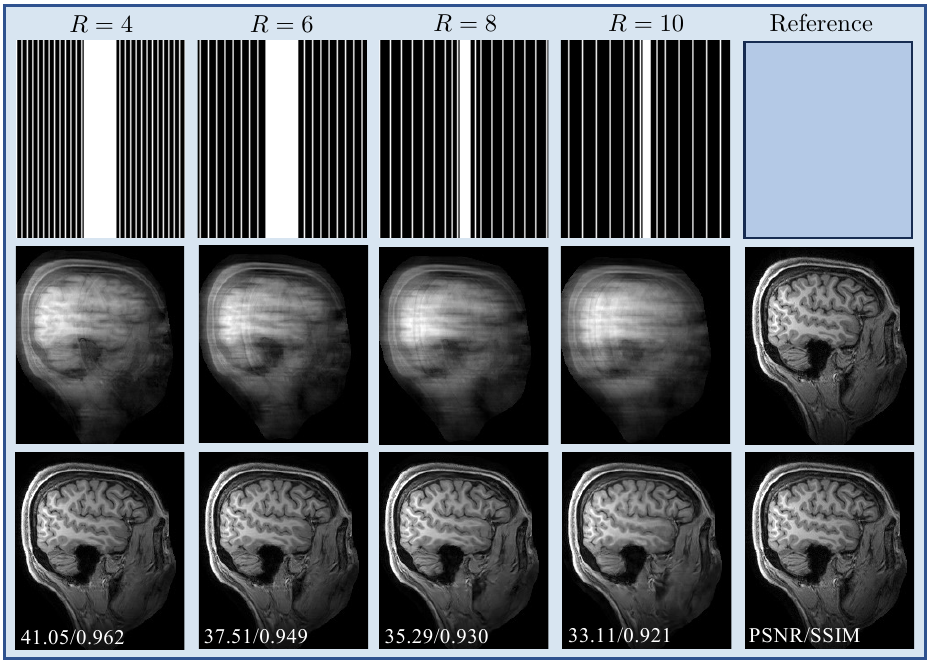} 
	\caption{The leftmost four in the first row shows the cartesian equispaced sampling masks $\Pbm$ used in the experimental validation: $\Rbm$ = $4$, $6$, $8$, $10$, with corresponding ACS lines = $24$, $24$, $8$ and $5$. The second row shows the zero-filled images of the same slice for the four acceleration rates. The third row shows the SPICER reconstructed images of the same slice and include the PSNR/SSIM values with respect to the reference as shown in the rightmost column.}
	\label{fig:masks}
\end{figure}

\subsection{In-vivo Brain Dataset}
The data acquisition was performed on a Siemens 3T Prisma scanner (Siemens Healthcare, Erlangen, Germany) with a 32-channel head coil. Images were collected with sagittal T1 magnetization-prepared rapid gradient-echo (MPRAGE) sequence. The acquisition parameters were as follows: repetition time (TR) = 2400 ms, echo time (TE) = 2.62 ms, inversion time (TI) = 1000 ms, flip angle (FA) = 8 degrees, FOV = 256 mm × 256 mm, voxel size = 1 × 1 × 1 mm, slices per slab = 176, slice and phase resolution = 100\% and slice and phase partial Fourier off. A $2 \times$ oversampling was used in the frequency encoding direction, and the asymmetric echo was turned on to allow short TE. Fully sampled measurements were acquired with GRAPPA turned off, and the total acquisition time was 10 minutes and 16 seconds. The reference images were obtained by the sum-of squares (SoS) reconstruction computed on the fully-sampled multi-coil data. Upon the approval of our Institutional Review Board, we used brain MRI data from 14, 1, and 5 participants in this study for training, validation, and testing, respectively. To obtain undersampled measurements, the multi-coil k-space data were retrospectively undersampled using 1D Cartesian equispaced sampling masks with ACS lines. The sampling strategy is based on the clinical-used $2\times$ acceleration sampling pattern, which is acquired with GRAPPA $\Rbm= 2$ in phase encoding (PE) direction with 24 ACS lines. 
As shown in Fig.~\ref{fig:masks}, we conducted our experiments for acceleration factors $\Rbm = 4$, $6$, $8$ and $10$ with 1D equispaced Cartesian masks, that contain $24$, $24$, $8$ and $5$ ACS lines respectively. They correspond to the retrospective sampling rates of $32\%$, $24\%$, $15\%$ and $12\%$. For each undersampling pattern, one pair of under-sampled measurements are generated for self-supervised training. We generate 1932 ($138 \times 14$) pairs of training data for each undersampling rate using the entire training dataset. Each training data pair corresponds to one of 138 slices from each of the 14 subjects in the dataset.

\subsection{fastMRI Brain Dataset}
The T2 MR brain dataset was obtained from the multi-coil fastMRI dataset. We used the T2 MR brain acquisitions of 165 subjects obtained from the multi-coil dataset as the raw reference. These 165 subjects were split into 130, 15, and 20 for training, validation, and testing, respectively. For each subject, we extracted the first 12 slices on the transverse plane, containing the most relevant regions of the brain. In order to obtain undersampled measurements, we use the 1D Cartesian equispaced sampling pattern provided in the fastMRI database with acceleration rate = 4, 8.

\subsection{Comparisons}
\subsubsection{Baseline Methods}
We selected several well-known methods as references to compare the performance of SPICER:
\begin{itemize}
	\item \emph{TV (with ESPIRiT)}: The traditional TV regularized image reconstruction~\cite{block2007undersampled}.
	\item \emph{GRAPPA}: Traditional GRAPPA~\cite{griswoldGeneralized2002} that linearly interpolates missing k-space points using nearby acquired k-space points from all coils.
	
	\item \emph{U-Net}: The U-Net architecture~\cite{ronnebergerUnet2015} trained to maximize SSIM between the reconstructed image and the image obtained using GRAPPA on the same amount of measurements as SPICER.
	
    \item \emph{SSDU}:  A well-known self-supervised learning method~\cite{yaman2020self} that trains a DU network by dividing each k-space MRI acquisition into two subsets and using them as training targets for each other. \markup{For fair comparison, SSDU network use the same U-Net architecture as SPICER.} 

    \item \emph{{${\text{SSDU}}^\text{auto}$}}: A variant of SSDU~\cite{yaman2020self} that incorporates a same automatic CSM learning module as SPICER. The strategy to generate training data pairs is the same with SSDU.
	
    \item \emph{{$\text{SSDU}^{*}$}}: A variant of SSDU~\cite{yaman2020self} that trained with paired training data and automatic CSM learning module as SPICER. The training data pairs are generated with Cartesian subsampling strategy (same as SPICER).
	
	\item \emph{E2E-VarNet}: An idealized variant of method from~\cite{sriramEndtoend2020} trained using SSIM on fully-sampled reference data. E2E-VarNet shows the upper bound on performance achievable by the self-supervised variants of the method.
\end{itemize} 

All the CSMs used in TV (with ESPIRiT), U-Net, and SSDU are pre-estimated using ESPIRiT~\cite{ueckerESPIRiTan2014}. We observed that when the number of ACS lines is less than 24, reconstruction using the CSMs estimated by ESPIRiT are of poor quality. Therefore, we use 24 ACS lines for estimating CSMs with ESPIRiT to be used in TV, U-Net, and SSDU.

\subsubsection{Ablation Study}

We also performed an ablation study to highlight the influence of the DMBA module, the CSMs estimation module, and the CSMs smoothness regularization within SPICER. We compared the original SPICER model with three different models as follows:

\begin{itemize}
\item \emph{Joint U-Net}: We replace the DMBA module of SPICER with U-Net to show the benefit of using a DMBA.
\item \emph{DMBA (ESPIRiT)}: We use the same DMBA as SPICER but use ESPIRiT on undersampled measurements to replace the CSM estimation module of SPICER.
\item \markup{\emph{{$\text{SSDU}^{*}$}}}: SPICER trained without the CSMs smoothness regularization loss $\mathrm{Loss}_\mathrm{smooth}$ in~\eqref{equ:SmoothingPrior}.
\end{itemize}

\subsubsection{Evaluation of Coil Sensitivity Maps Estimation}
In order to evaluate the quality of the estimated CSMs, we use CSMs estimated by the proposed and ESPIRiT within the iterative TV reconstruction. We use the \texttt{\small fminbound} method in the \texttt{\small scipy.optimize toolbox} to find the optimal regularization parameters for TV.

\begin{figure}[H]
\centering \includegraphics[width=.975\textwidth]{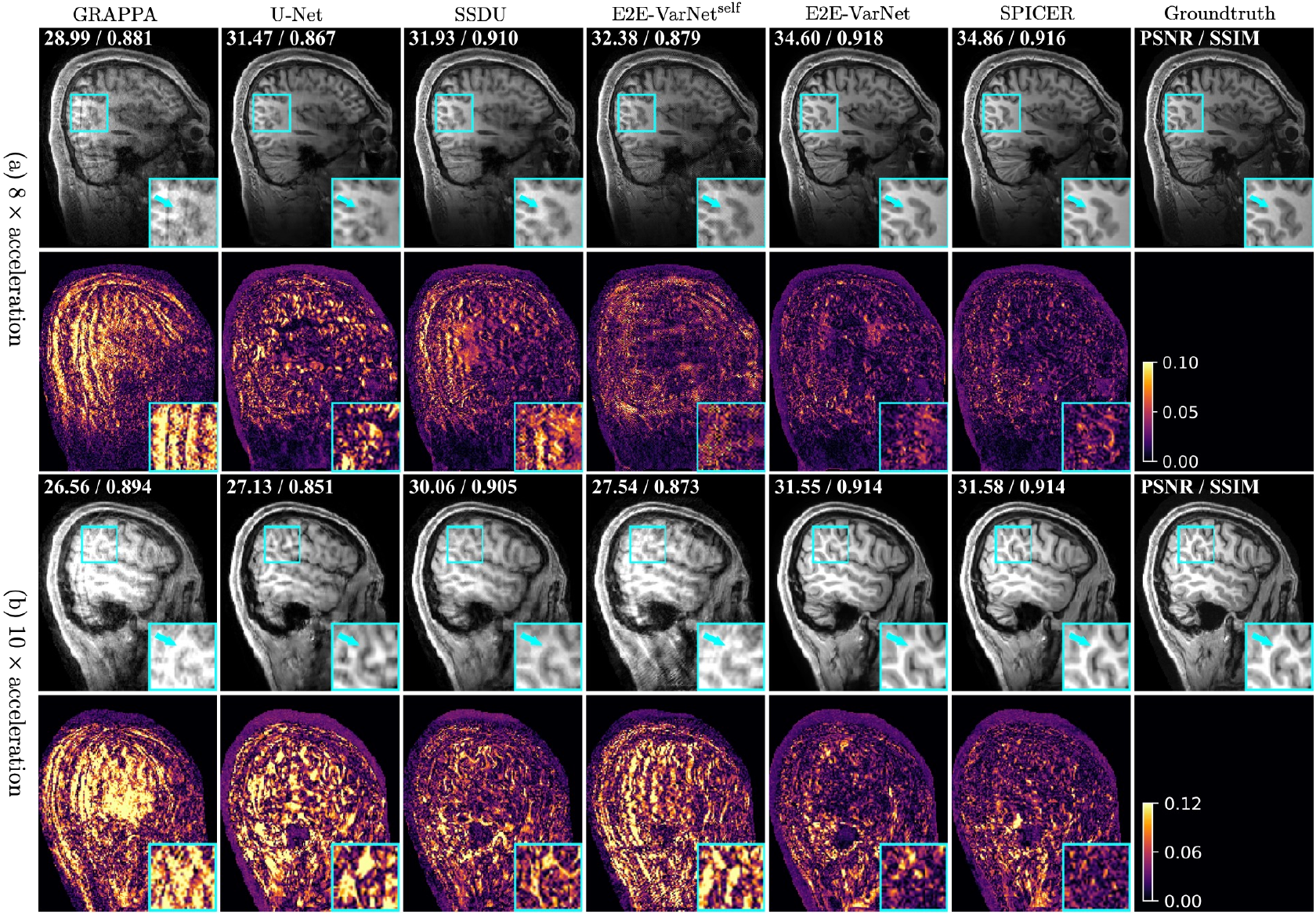} 
	\caption{Visual and quantitative evaluation on in-vivo brain dataset corresponding to $8\times$ and $10\times$ acceleration rates. The top-right corner of each image provides the PSNR and SSIM values with respect to the reference. We highlight the visually significant differences using zoom views and error maps. The visually important differences are highlighted using arrows. U-Net, $\text{SSDU}^\text{auto}$, and SPICER in the figure are based on self-supervised learning.
	$\text{E2E-VarNet}$ is a supervised learning method trained using groundtruth. SPICER achieves the best performance compared to the baseline methods by jointly performing image reconstruction and CSM estimation with end-to-end self-supervised training. Note how compared to other methods, SPICER recovers sharper images and reduces artifacts, even achieving comparable performance to the supervised learning method.}
	\label{fig:exp-img}
\end{figure}

\subsection{ Evaluation Metrics}
For in-vivo Brain Dataset, SPICER was compared  with all the baselines shown in section 3.8 at acceleration rates R = 4, 6, 8, and 10. For fastMRI Brain Dataset,  SPICER  was  compared  with  TV, GRAPPA, SSDU, and supervised method E2E-VarNet trained on fully sampled data at  the  same  acceleration  R  =  4 and 8. We used widely-used quantitative metrics, peak signal-to-noise ratio (PSNR), measured in dB, structural similarity index (SSIM), and the normalized mean squared error (NMSE), with respective to the reference images obtained from the fully-sampled data. 
The quantitative results were statistically analyzed by comparing SPICER with other image reconstruction methods within the targeted brain region, using the same threshold operator in ESPIRiT. We used the non-parametric Friedman’s test and the post-hoc test of the original FDR method of Benjamini
and Hochberg~\cite{benjamini1995controlling}. The statistical analysis was performed using GraphPad Prism 9 (Version 9.3.1 for macOS, GraphPad
Software, San Diego, CA, USA). Statistical significance was
defined as $P$ \textless $0.05$.

\begin{figure}[H]
\centering \includegraphics[width=.975\textwidth]{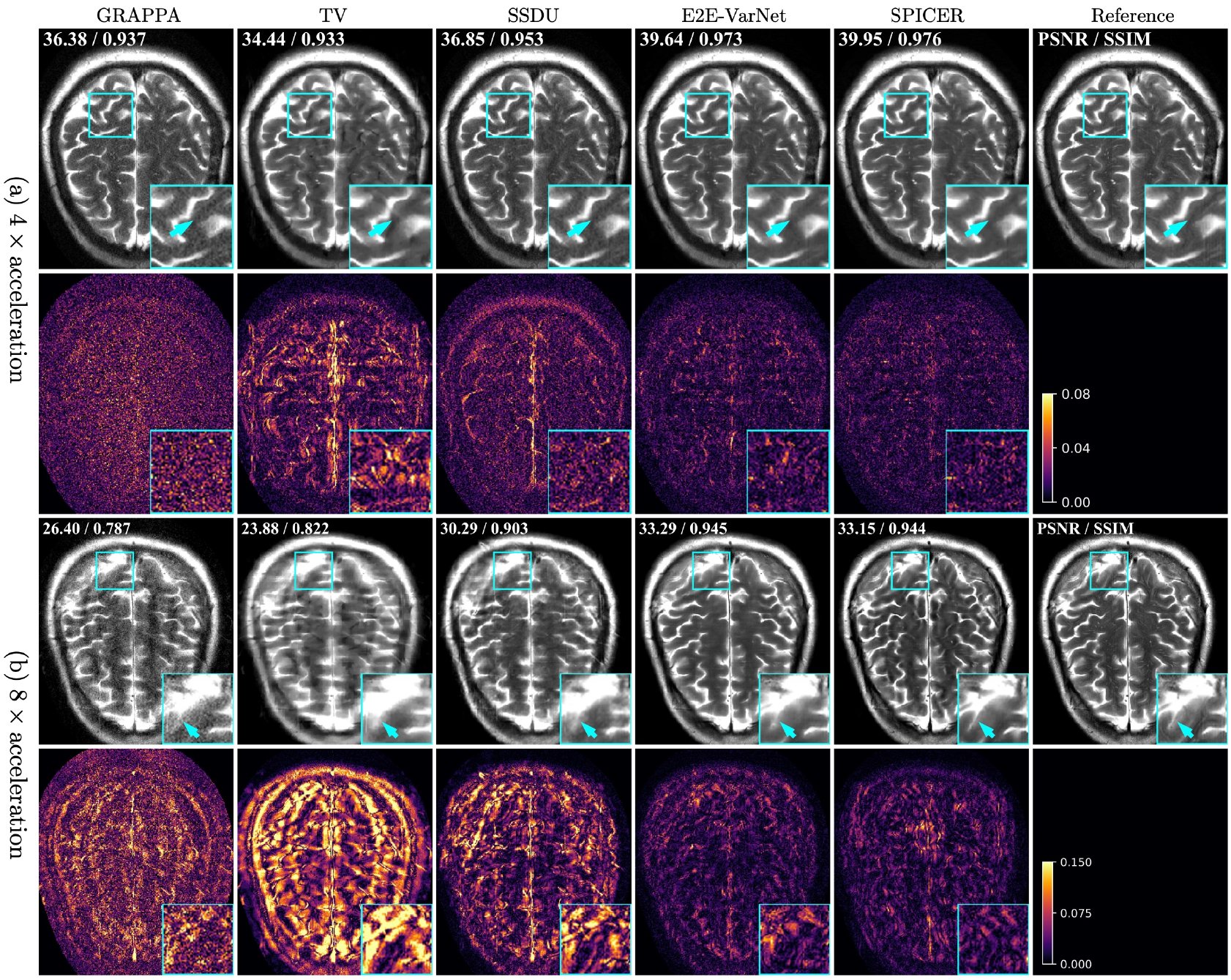} 
	\caption{Visual and quantitative evaluation on fastmri brain dataset corresponding to $4\times$ and $8\times$ acceleration rates. The top-right corner of each image provides the PSNR and SSIM values with respect to the reference. We highlight the visually significant differences using zoom views and error maps. The visually important differences are highlighted using arrows. SSDU and SPICER in the figure are based on self-supervised learning.
	$\text{E2E-VarNet}$ is a supervised learning method trained using groundtruth. SPICER achieves the best performance compared to the self-supervised baseline methods by jointly performing image reconstruction and CSM estimation with end-to-end self-supervised training. Note how compared to other methods, SPICER recovers sharper images and reduces artifacts, even achieving comparable performance to the supervised learning method.}
	\label{fig:exp-img-fastmri}
\end{figure}

\section{Results}
\label{sec:results}

We now present the numerical validation of SPICER on both \emph{in-vivo} MRI dataset and fastMRI dataset. The results show that SPICER achieves state-of-the-art performance in highly accelerated acquisition settings. For better visualization, the 2D visualization cases in Figure 3-6 are chosen randomly from middle 50 slides of the 3D volume.

Fig.~\ref{fig:exp-img} illustrates the results of image reconstruction on the in-vivo dataset for acceleration rates $\Rbm=8$ (top) and $\Rbm=10$ (bottom). TV (with ESPIRiT) and GRAPPA contain blurring and ghosting artifacts, especially at higher acceleration rates. While U-Net has better performance than TV (with ESPIRiT) and GRAPPA by learning the prior from data, SSDU and $\text{SSDU}^\text{auto}$ outperform it due to their model-based DL architectures.
$\text{SSDU}^\text{auto}$ performs joint CSMs estimation, which enables it to do better than SSDU. \markup{However, it continues to exhibit blurring, stemming from the incongruence between the training and testing phases, further intensified by the absence of paired data utilization. This issue arises due to the undersampled measurement splitting strategy employed, highlighting the need for paired undersampled data and corresponding new splitting strategy.} The supervised learning baseline E2E-VarNet and our proposed method SPICER achieve significant improvements compared to other methods. Overall, our proposed method SPICER can provide the best performance in artifact removal and sharpness compared to all of the self-supervised baseline methods.

Table~\ref{tb:main} summarizes quantitative results of all
the evaluated methods. In Table~\ref{tb:main}, SPICER achieves the highest PSNR, highest SSIM, and lowest NMSE values compared to other methods over all considered acceleration rates. Notably, SPICER outperformed all variants of the SSDU method, which highlights that the success of SPICER comes not merely from the access to paired
undersampled data or a simplistic automated CSM estimation module. Moreover, our method achieves very competitive performance with a state-of-the-art supervised learning method E2E-VarNet. Note that E2E-VarNet also utilizes the joint learning of unrolling network and CSM estimation network. Statistical analysis of PSNR, SSIM, and NMSE values in Table~\ref{tb:main} also highlights that SPICER can achieve statistically significant results compared to all the self-supervised image reconstruction baseline methods.

Figure~\ref{fig:exp-img-fastmri} shows the results of image reconstruction on fastMRI dataset for acceleration rates $\Rbm=4$ (top) and $\Rbm=8$ (bottom). TV (with ESPIRiT) and GRAPPA contain artifacts and noise, especially at higher acceleration rates. SSDU successfully removes most of the artifacts and noise, but it still suffers from some blurring, due to the use of sub-optimal pre-calibrated CSMs. The supervised learning baseline E2E-VarNet and our proposed self-supervised method SPICER significantly outperform other methods. Overall, our proposed method SPICER can provide the best performance in artifact removal and sharpness compared to all of the self-supervised baseline methods on fastMRI dataset.

Table~\ref{tb:main_fastmri} presents the quantitative analysis of all the evaluated methods on fastMRI dataset. Table~\ref{tb:main_fastmri} shows that SPICER achieves the highest PSNR, highest SSIM, and lowest NMSE values compared to other self-supervised methods over all considered acceleration rates. Moreover, our method achieves competitive performance with the supervised learning method E2E-VarNet.

\begin{figure}
	\centering \includegraphics[width=.945\textwidth]{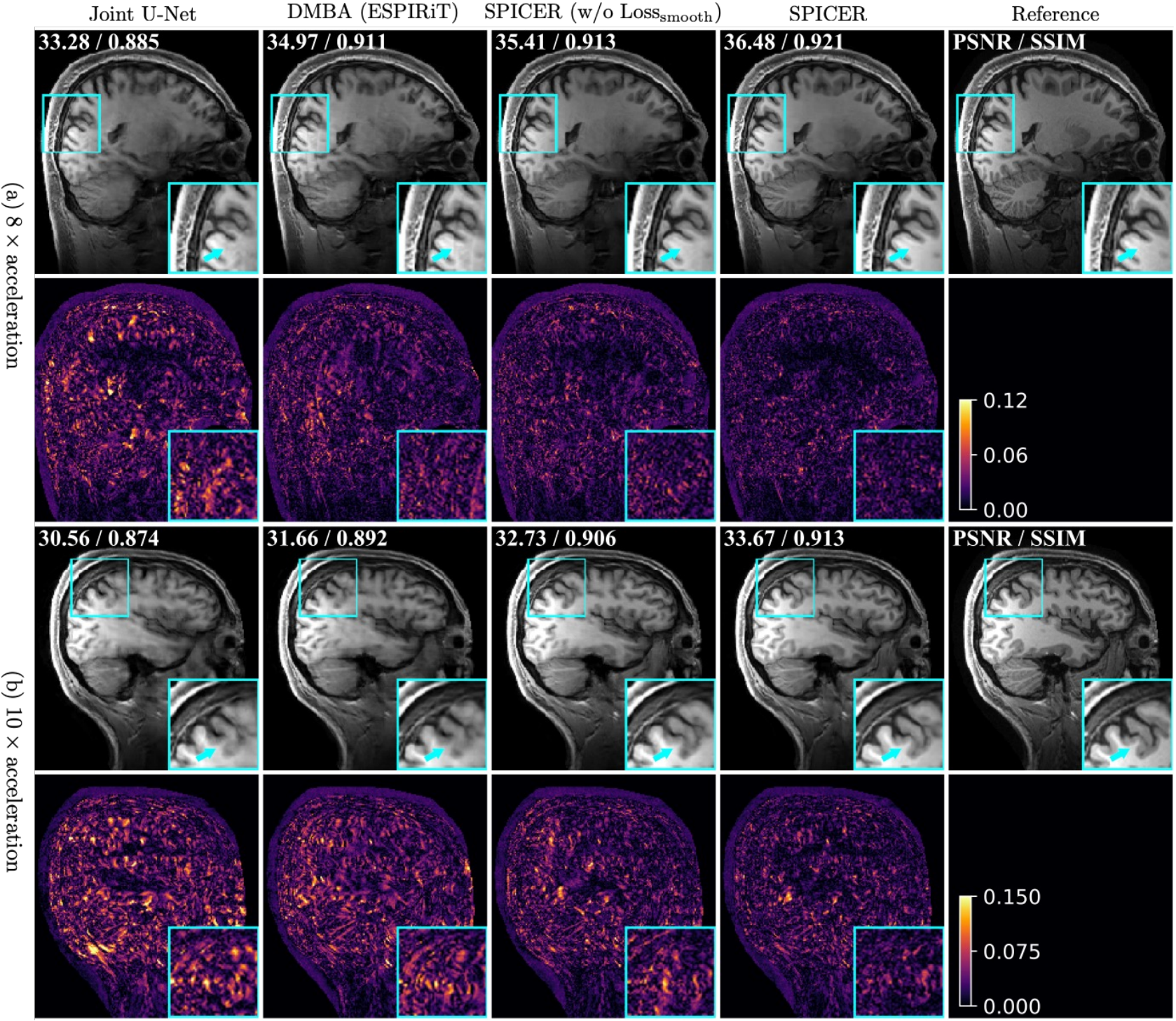} 
	\caption{Quantitative evaluation of SPICER on in-vivo brain dataset at $8\times$ and $10\times$ acceleration rates. The top-right corner of each image provides the PSNR and SSIM values with respect to the reference. We highlight visually significant differences using zoom views and error maps. This figure shows that the DMBA, CSM estimator, and CSM smoothness regularization of SPICER play important role in enhancing both quantitative and visual performance in terms of both artifact-removal and sharpness.}
	\label{fig:ablation}
\end{figure}

Figure~\ref{fig:ablation} illustrates reconstruction results of the ablation study on $8\times$ and $10\times$ acceleration. Figure~\ref{fig:ablation} shows that the DMBA model, CSM estimator, and CSM smoothness regularization of SPICER improve the imaging quality. 
Table~\ref{tb:ablation} presents the results of an ablation study showing the influence of each module of our SPICER model and the effectiveness of $\mathrm{Loss}_\mathrm{smooth}$ in the proposed self-supervised loss function \eqref{equ:loss}. As can be seen, Joint U-Net obtains the worst performance, due to the U-Net model not using the PMRI measurement model. The results suggest that pre-estimated CSMs generated with ESPIRiT used in DMBA lead to sub-optimal performance in all acceleration rates, especially with limited ACS lines. Note that at high acceleration rates ($8\times$ and $10\times$), even the CSMs pre-estimated with ESPIRiT on $4\times$ accelerated data, the performance is still much worse than that of the joint CSM learning methods. The results of \markup{\emph{{$\text{SSDU}^{*}$}}} and SPICER show that CSM smoothness regularization plays an important role in the reconstruction. Statistical analysis of quantitative
values in Table~\ref{tb:ablation} highlights that SPICER achieves statistically significant results compared to other methods.

\begin{figure}
	\centering \includegraphics[width=.935\textwidth]{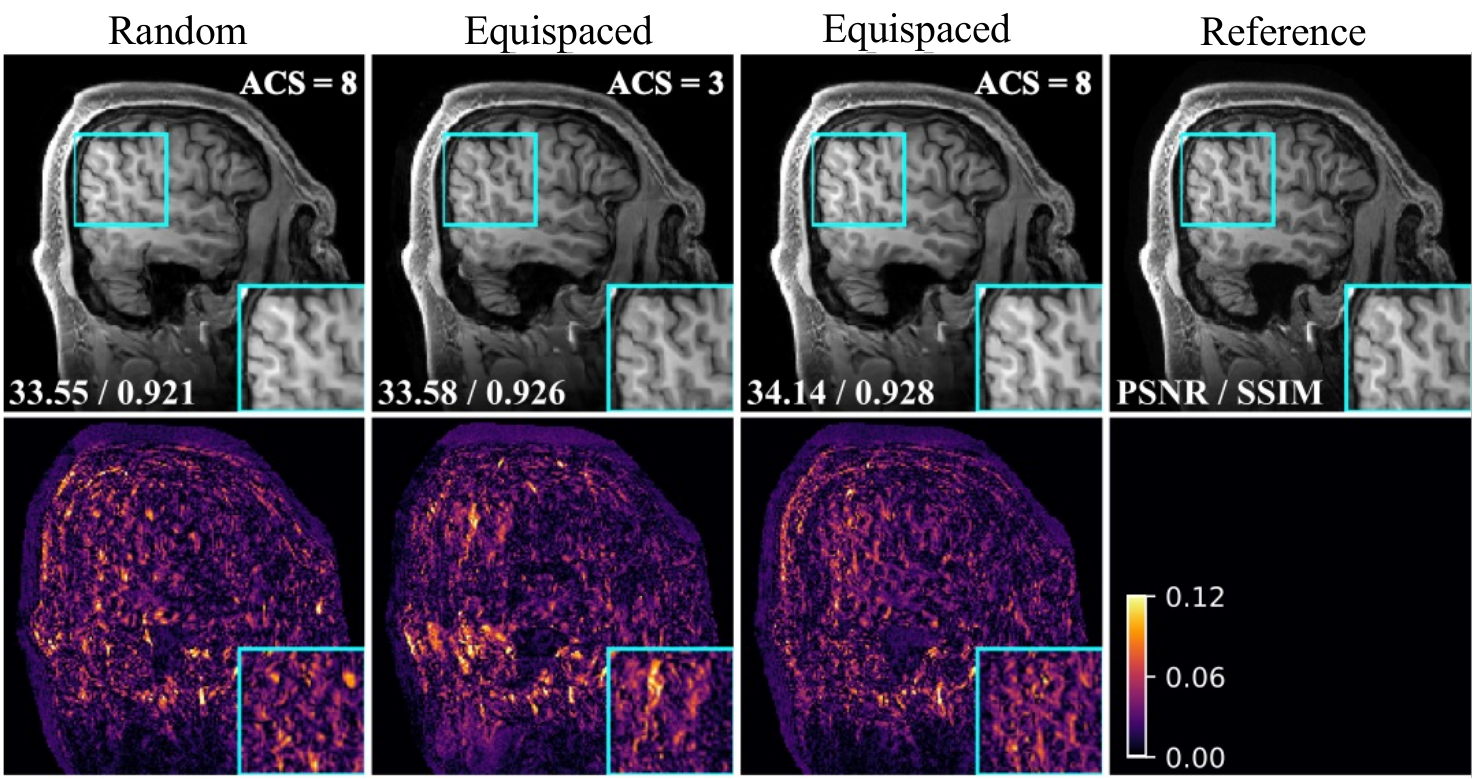} 
	\caption{Visual and quantitative evaluation on in-vivo brain dataset corresponding to different sampling strategies for $8\times$ acceleration rate. The first row shows the results from 1D random sampling, 1D equispaced sampling with only 3 ACS lines, 1D equispaced sampling with 8 ACS lines and reference. The bottom-left corner of each image provides the PSNR and SSIM values with respect to the reference. The top-right corner of each image provides the number of ACS lines used for each undersampling pattern.  Visually significant differences are highlighted using error maps.}
	\label{fig:different_sample}
\end{figure}

Fig.~\ref{fig:different_sample} illustrates the performance of our SPICER method using various sampling patterns. We consider three distinct sampling patterns: 1D random sampling with 8 ACS lines, 1D equispaced sampling with 3 ACS lines, and 1D equispaced sampling with 8 ACS lines. The results suggest that SPICER can consistently provide high-quality reconstruction results even even in challenging scenarios involving random sampling patterns and extremely limited ACS regions (only three lines).

\begin{figure}
    \centering 	\includegraphics[width=.935\textwidth]{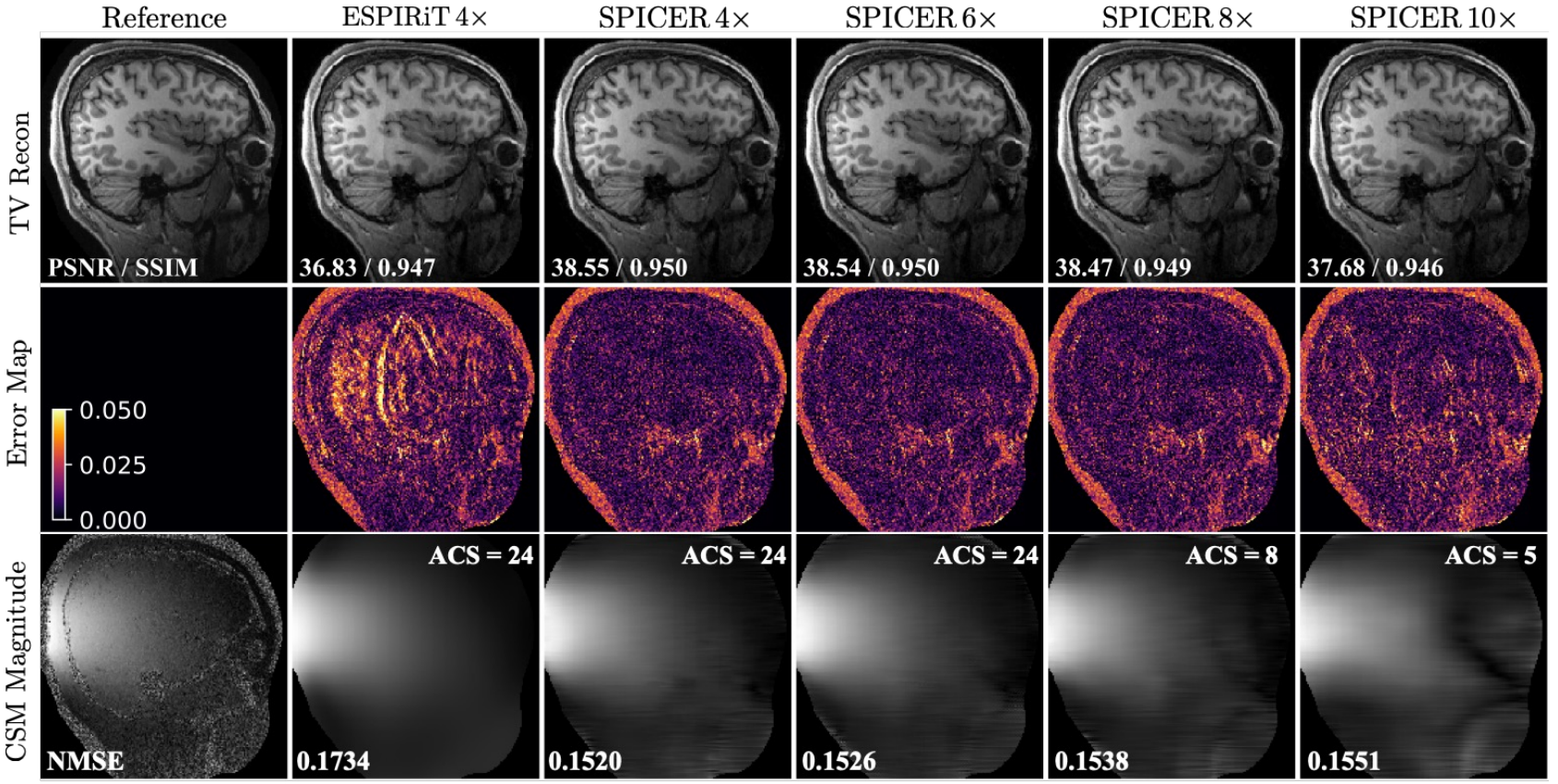}
	\caption{Evaluation of CSMs estimated by SPICER at various acceleration rates. The first row shows the TV reconstruction results at acceleration rate $\Rbm=4$ with the corresponding CSM shown in the third row. The bottom-left corner of each reconstruction image provides the PSNR and SSIM values with respect to the reference. The third row shows the magnitude image of the 10th CSM. For each undersampling pattern, the number of ACS lines used is indicated in the top-right corner of each CSM image. The bottom-left corner of each CSM image includes the NMSE value relative to the reference CSM.  Visually significant differences are highlighted using error maps in the second row.}
	\label{fig:csm_tv}
\end{figure}

Fig.~\ref{fig:csm_tv} illustrates the performance of the estimated CSMs. We use ESPIRiT and SPICER to estimate CSMs from data at various accelerated acquisition rates (from $4\times$ to $10\times$). Estimated CSMs are then used within TV reconstruction from data corresponding to $4\times$ accelerated acquisition rate. The reconstructed images are quantitatively evaluated using PSNR and SSIM values relative to the reference images. The top row of Fig.~\ref{fig:csm_tv} illustrates the TV reconstruction results equipped with CSMs estimated from different ACS data and different methods. The bottom row of Fig.~\ref{fig:csm_tv} illustrates the visualization of CSMs of one specific coil. Note that, even with CSMs estimated by SPICER from 5 ACS lines, the following TV reconstruction can achieve better performance than ESPIRiT CSMs (generated from 24 ACS lines).

\begin{table}[hbt!]
\centering
\footnotesize
	\begin{threeparttable}
     \renewcommand\arraystretch{1.35}
		\setlength\tabcolsep{2pt}

		\begin{tabular}{c|ccc||ccc}
		\hline
			\rowcolor{lightgray}	\emph{Acceleration Factor}                   & \multicolumn{3}{c}{$4\times$ acceleration }  & \multicolumn{3}{c}{$6\times$ acceleration }\\
			\rowcolor{lightgray}	\emph{Method}             & \multicolumn{1}{c}{PSNR $\uparrow$} & \multicolumn{1}{c}{SSIM $\uparrow$ }   &\multicolumn{1}{c}{NMSE $\downarrow$}  & \multicolumn{1}{c}{PSNR $\uparrow$} & \multicolumn{1}{c}{SSIM $\uparrow$ }   &\multicolumn{1}{c}{NMSE $\downarrow$ } \\ \hline
			{Zero-Filled}                &$18.27 \pm 1.672 $$\star$ &$ 0.775  \pm 0.3842 $$\star$ &$ 0.6212  \pm 0.3460 $$\star$  &$18.07 \pm 1.694 $$\star$ &$ 0.745  \pm 0.0419$$\star$ &$ 0.6532  \pm 0.3732 $$\star$ \\
			{Total variation~\cite{block2007undersampled}}  & $37.87\pm1.491$$\star$ &$0.959\pm0.0097$$\star$ &$0.0043\pm0.0019$$\star$ & $32.99\pm1.305$$\star$ &$0.898\pm0.0152$$\star$ &$0.0171\pm0.0050$$\star$\\
		{	GRAPPA~\cite{griswoldGeneralized2002}} &$38.54\pm1.467$$\star$ &$0.966\pm0.0071$$\star$ &$0.0039\pm0.0012$$\star$ & $33.05\pm1.205$$\star$ &$0.902\pm0.0151$$\star$ &$0.0471\pm0.0132$$\star$ \\
			{	U-Net~\cite{ronnebergerUnet2015}} & $35.36\pm1.45 
  5$$\star$ &$0.925\pm0.0139$$\star$ & $0.0111\pm0.0034$$\star$ & $33.33\pm1.425$$\star$ &$0.903\pm0.0160$$\star$& $0.0179\pm0.0051$$\star$\\
			{	SSDU~\cite{yaman2020self}} & $36.52\pm1.676$$\star$ & $0.942\pm0.0116$$\star$  & $0.0086\pm0.0048$$\star$ & $33.91\pm2.021$$\star$ & $0.921\pm0.0144$$\star$  & $0.0176\pm0.0191$$\star$\\
		{$\text{SSDU}^\text{auto}$}  & $37.07\pm1.485$$\star$ & ${0.943}\pm0.0125$$\star$  & $ 0.0074\pm0.0041$$\star$  & $33.24\pm1.520$$\star$ &$0.912\pm0.0181$$\star$& $0.0182\pm0.0082$$\star$\\

            $\text{SSDU}^{*}$  & $38.21\pm1.637$$\star$ &$0.945\pm0.0129$$\star$ & $0.0050\pm0.0019$$\star$ & $36.19\pm1.499$$\star$ &$0.929\pm0.01307$$\star$ & $0.0095\pm0.0040$$\star$\\

           {SPICER} &$\bf{38.93}\pm1.620$ &$\bf0.949\pm0.0111$ & $\bf{0.0048}\pm0.0017$ &$\bf{36.77}\pm1.575$ &$\bf{0.935}\pm0.0120$ & $\bf{0.0081}\pm0.0037$\\
			
			\cdashline{1-7}
			{E2E-VarNet~\cite{sriramEndtoend2020}}  & $38.56\pm1.768$$\star$ & $0.952\pm0.0113$$\star$  & $0.0054\pm0.0030$$\star$  & $36.09\pm1.491$$\star$ & $0.938\pm0.0136$$\star$ & $0.0095\pm0.0056$$\star$\\
	\hline
		
	\end{tabular}
    \vspace*{0.15cm}
	\begin{tabular}{c|ccc||ccc}
	\hline
		\rowcolor{lightgray}	\emph{Acceleration Factor}                   & \multicolumn{3}{c}{$8\times$ acceleration }  & \multicolumn{3}{c}{$10\times$ acceleration }\\
		\rowcolor{lightgray}	\emph{Method}             & \multicolumn{1}{c}{PSNR $\uparrow$} & \multicolumn{1}{c}{SSIM $\uparrow$ }   &\multicolumn{1}{c}{NMSE $\downarrow$}  & \multicolumn{1}{c}{PSNR $\uparrow$} & \multicolumn{1}{c}{SSIM $\uparrow$ }   &\multicolumn{1}{c}{NMSE $\downarrow$ } \\ \hline
		{Zero-Filled}                &$17.55 \pm 1.510 $$\star$ &$ 0.721  \pm 0.0433 $$\star$ &$ 0.7197  \pm 0.3737 $$\star$ & $16.90 \pm 1.328 $$\star$ &$ 0.701  \pm 0.0450 $$\star$ &$ 0.8209  \pm 0.3865 $$\star$\\
		{Total variation~\cite{block2007undersampled}}  & $26.09\pm1.218$$\star$ &$0.878\pm0.0157$$\star$ &$0.1062\pm0.0295$$\star$ & $25.28\pm1.487$$\star$ &$0.863\pm0.0156$$\star$ &$0.1331\pm0.0519$$\star$\\
		{GRAPPA~\cite{griswoldGeneralized2002}} & $28.31\pm1.390$$\star$  &$0.892\pm0.0205$$\star$ &$0.0568\pm0.0221$$\star$ & $25.79\pm1.177$$\star$ &$0.873\pm0.0185$$\star$ &$0.0924\pm0.0354$$\star$\\
		{	U-Net~\cite{ronnebergerUnet2015}} & $30.81\pm1.543$$\star$ &$0.868\pm0.0218$$\star$ & $0.0333\pm0.0122$$\star$ & $29.17\pm1.495$$\star$ &$0.836\pm0.0256$$\star$& $0.0489\pm0.0207$$\star$\\
		{	SSDU~\cite{yaman2020self}} & $32.58\pm1.534$$\star$ & $0.910\pm0.0154$$\star$ & $0.0216\pm0.0127$$\star$ & $31.08\pm1.567$$\star$ & $0.892\pm0.0185$$\star$ & $0.0310\pm0.0218$$\star$ \\ 
	{$\text{SSDU}^\text{auto}$}  & $31.30\pm1.634$$\star$ & ${0.880}\pm0.0198$$\star$  & $0.0281\pm0.0112$$\star$  & $30.45\pm1.463$$\star$ &$0.878\pm0.0204$$\star$& $0.0347\pm0.0198$$\star$\\

 	$\text{SSDU}^{*}$ & $34.26\pm1.539$$\star$ &$0.910\pm0.0169$$\star$ & $0.0165\pm0.0049$$\star$ & $32.14\pm1.386$$\star$ &$0.894\pm0.0208$$\star$& $0.0228\pm0.0070$$\star$\\

		{SPICER} &$\bf{34.66}\pm1.397$ &$\bf{0.918}\pm0.0153$ & $\bf{0.0128}\pm0.0036$ &$\bf{32.94}\pm1.456$ &$\bf{0.909}\pm0.0191$ & $\bf{0.0176}\pm0.0067$\\
		
		\cdashline{1-7}
		{E2E-VarNet~\cite{sriramEndtoend2020}}  & $34.18\pm1.487$$\star$ & ${0.920}\pm0.168$$\star$  & $0.0144\pm0.0059$$\star$  &
  $32.91\pm1.576$$\star$ & $0.919\pm0.0141$$\star$  & $0.0194\pm0.0083$$\star$\\
	    \hline
	\end{tabular}
	\begin{tablenotes}
		\item Statistically significant differences compared with SPICER are marked ($P\star$ \textless $0.05$).
	\end{tablenotes}
	\end{threeparttable}
	\caption{Quantitative evaluation (Mean $\pm$ Standard Deviation) of SPICER. Note how SPICER achieves the best performance against other self-supervised baselines and comparable performance relative to the supervised method.}
	\label{tb:main}
\end{table}

\begin{table}[hbt!]
  \centering
  \footnotesize
	\begin{threeparttable}
		\renewcommand\arraystretch{1.35}
		\setlength\tabcolsep{2pt}
				\begin{tabular}{c|ccc||ccc}
			\hline
			
			\rowcolor{lightgray}\emph{Acceleration Factor}                   & \multicolumn{3}{c}{$4\times$ acceleration }  & \multicolumn{3}{c}{$6\times$ acceleration }\\
			\rowcolor{lightgray}\emph{Method}             & \multicolumn{1}{c}{PSNR $\uparrow$} & \multicolumn{1}{c}{SSIM $\uparrow$ }   &\multicolumn{1}{c}{NMSE $\downarrow$}  & \multicolumn{1}{c}{PSNR $\uparrow$} & \multicolumn{1}{c}{SSIM $\uparrow$ }   &\multicolumn{1}{c}{NMSE $\downarrow$ } \\ \hline
			{Joint U-Net} &$35.77\pm 1.352 $$\star$ &$ 0.950  \pm 0.0119 $$\star$ &$ 0.0105  \pm 0.0030 $$\star$ & $33.45 \pm 1.525 $$\star$ &$ 0.911 \pm 0.0179 $$\star$ &$ 0.0168  \pm0.0054 $$\star$\\
			{DMBA (ESPIRiT)}   & $36.80\pm 1.502 $$\star$ &$ 0.959  \pm 0.0098 $$\star$ &$ 0.0079  \pm 0.0024 $$\star$  & $35.21\pm1.313$$\star$ &$0.923\pm0.01557$$\star$ & $0.0111\pm0.0032$$\star$ \\
			\markup{{$\text{SSDU}^{*}$}} & $38.21\pm1.637$$\star$ &$0.945\pm0.01299$$\star$ & $0.0050\pm0.0019$$\star$ & $36.19\pm1.499$$\star$ &$0.929\pm0.01307$$\star$ & $0.0095\pm0.0040$$\star$\\
         {SPICER} &$\bf{38.93}\pm1.620$ &$\bf0.949\pm0.0111$ & $\bf{0.0048}\pm0.0017$ &$\bf{36.77}\pm1.575$ &$\bf{0.935}\pm0.0120$ & $\bf{0.0081}\pm0.0037$\\
			\hline
	
		\end{tabular}
		
    \vspace*{0.15cm}
	
		
	\begin{tablenotes}
		\item Statistically significant differences compared with SPICER are marked ($P\star$ \textless $0.05$).
	\end{tablenotes}
	\end{threeparttable}
	\caption{Quantitative evaluation (Mean $\pm$ standard deviation) using four distinct implementations of SPICER. This table shows the influence of the DMBA reconstruction module, the CSM estimation module, and the smooth term in the loss function.}
	\label{tb:ablation}
\end{table}

\begin{table}[hbt!]
\centering
\footnotesize
	\begin{threeparttable}
     \renewcommand\arraystretch{1.35}
		\setlength\tabcolsep{2pt}
		\begin{tabular}{c|ccc||ccc}
		\hline
			\rowcolor{lightgray}	\emph{Acceleration Factor}                   & \multicolumn{3}{c}{$4\times$ acceleration }  & \multicolumn{3}{c}{$8\times$ acceleration }\\
			\rowcolor{lightgray}	\emph{Method}             & \multicolumn{1}{c}{PSNR $\uparrow$} & \multicolumn{1}{c}{SSIM $\uparrow$ }   &\multicolumn{1}{c}{NMSE $\downarrow$}  & \multicolumn{1}{c}{PSNR $\uparrow$} & \multicolumn{1}{c}{SSIM $\uparrow$ }   &\multicolumn{1}{c}{NMSE $\downarrow$ } \\ \hline
			{Zero-Filled}                &$23.61 \pm 1.938 $$\star$ &$ 0.774  \pm 0.0417 $$\star$ &$ 0.0701  \pm 0.0445 $$\star$  &$18.51 \pm 1.898 $$\star$ &$ 0.601  \pm 0.0509$$\star$ &$ 0.244  \pm 0.1612 $$\star$ \\
			{Total variation~\cite{block2007undersampled}}  & $33.34\pm1.477$$\star$ &$0.939\pm0.0092$$\star$ &$0.0128\pm0.0070$$\star$ & $24.86\pm1.651$$\star$ &$0.819\pm0.0358$$\star$ &$0.0935\pm0.0570$$\star$\\
			{GRAPPA~\cite{griswoldGeneralized2002}} &$35.69\pm1.022$$\star$  &$0.893\pm0.0290$$\star$ &$0.0071\pm0.0025$$\star$ & $26.71\pm1.292$$\star$ &$0.724\pm0.0661$$\star$ &$0.0584\pm0.0292$$\star$ \\
			{SSDU~\cite{yaman2020self}} & $38.79\pm1.052$$\star$ & $0.953\pm0.0104$$\star$  & $0.0034\pm0.0012$$\star$ & $27.63\pm2.135$$\star$ &$0.876\pm0.0213$$\star$ &$0.0489\pm0.0319$$\star$\\
		
           {SPICER} &$\bf{39.64}\pm0.976$ &$\bf{0.970}\pm0.0051$ & $\bf{0.0028}\pm0.0011$ &$\bf{33.48}\pm1.648$ &$\bf0.946\pm0.0094$ & $\bf{0.0116}\pm0.0042$\\
			
			\cdashline{1-7}
			{E2E-VarNet~\cite{sriramEndtoend2020}}  & $39.46\pm1.128$$\star$ & $0.972\pm0.0063$$\star$  & $0.0030\pm0.0014$$\star$  & $34.70\pm1.341$$\star$ & $0.950\pm0.0084$$\star$ & $0.0093\pm0.0054$$\star$\\
	\hline
		
	\end{tabular}

	\begin{tablenotes}
		\item Statistically significant differences compared with SPICER are marked ($P\star$ \textless $0.05$).
	\end{tablenotes}
	\end{threeparttable}
	\caption{Quantitative evaluation (Mean $\pm$ Standard Deviation) of SPICER on fastMRI dataset. Note how SPICER achieves the best performance against other self-supervised baselines and comparable performance relative to the supervised method.}
	\label{tb:main_fastmri}
\end{table}

\clearpage
\section{Discussion and Conclusions}
\label{sec:conclusion}
\subsection{Discussion}
In this manuscript, we proposed a self-supervised DMBA, namely SPICER, for joint MRI reconstruction and automatic coil sensitivity calibration. The key benefit of SPICER is that it is trained directly on pairs of noisy and undersampled k-space measurements of the same object without any fully-sampled groundtruth data, which makes it broadly applicable when groundtruth is impossible or difficult to obtain. In addition, the automatic CSMs estimation module in SPICER shows the potential to estimate the high quality from only a few ACS lines, which can sharply increase the acceleration rate for the Cartesian sampling strategy.

Despite being trained on undersampled and noisy data, numerical results on the in-vivo Brain dataset demonstrate that SPICER can significantly outperform every self-supervised baseline while still achieving comparable reconstruction performance with the widely-used supervised method E2E-VarNet. Additionally, we further validated SPICER's efficacy using data from fastMRI dataset. Both qualitatively and quantitatively, the reconstruction performances of SPICER at the acceleration rates of 4, 6, 8, and 10 consistently surpassed the state-of-the-art self-supervised method SSDU.

The proposed SPICER model achieves comparable results to the supervised learning method E2E-VarNet, primarily due to our training loss function~\eqref{equ:unfolded} being specified in the k-space domain, which exhibits greater robustness to measurement noise. The training approach employed by E2E-VarNet relies on the SSIM loss specified on image domain, which makes it sensitive to noise and artifacts in the training data due to image formation. Consequently, SPICER demonstrates superior PSNR values compared to E2E-VarNet, albeit with slightly lower SSIM values.

We performed an ablation study to highlight the influence of the DMBA module, the CSMs estimation module, and the CSMs smoothness regularization within SPICER. As shown in Figure~\ref{fig:ablation} and Table~\ref{tb:ablation}, the SPICER DMBA can lead to a sharp performance improvement compared with traditional U-Net model, due to the U-Net model not using the PMRI measurement model. In addition, the joint estimation module shows the potential to estimate more accurate CSM compared to the widely-used ESPIRiT method, which enables SPICER to achieve consistently better performance in all acceleration rates, especially with
limited ACS lines. Additionally, the SPICER loss function performs better both qualitatively and statistically with the benefit of the CSM smoothness regularization.

We performed comparison to several variants of SSDU to emphasize that the superior performance of SPICER extends beyond our paired training data generation strategy. Table~\ref{tb:main} illustrates that directly incorporating established self-supervised training techniques and an automatic CSM learning module resulted in minimal or negligible improvements. Notably, even when $\text{SSDU}^{*}$ employs the same training data splitting methodology as SPICER to generate data pairs, it demonstrates sub-optimal performance. This performance gap arises from SPICER's unique training approach, which not only involves the utilization of Cartesian-subsampled data pairs but also incorporates CSMs with smoothness regularization. This integration plays a pivotal role in achieving the superior performance observed in comparison to $\text{SSDU}^{*}$.

\subsection{Limitations and future directions}
Several limitations exist for our proposed deep learning reconstruction framework. First, our numerical study is currently limited to brain images. However, the approach of jointly using self-supervised learning for coil sensitivity maps and DMBA for accurate image reconstruction can be extended to a variety of imaging scenarios, provided that high-quality training data is available. Future research could explore the adaptation of this approach to other body parts, particularly in cases where body coils lack a fixed geometry, which could pose additional challenges.

We furthermore recognize that our model's current training and testing are limited to Cartesian sampling trajectories. There is growing interest in extending SPICER to non-Cartesian sampling trajectories, such as spiral and radial trajectories. These trajectories offer the advantage of densely sampling the central k-space region, and the resulting zero-filled images can be used as inputs to the CSM estimation module, providing similar benefits to the ACS lines in Cartesian sampling. The extension of SPICER to non-Cartesian sampling presents an exciting opportunity to improve the technique’s performance and applicability in various imaging scenarios that demand high spatiotemporal resolution or improved motion robustness.

Another limitation of SPICER is that it performs 3D reconstruction slice-by-slice. To enhance its capabilities, we envision the formulation of a 3D variant for reconstructing 3D volumes in cases where undersampling may occur in two dimensions. Such a development can potentitally further accelerate the MRI acquisition.

\markup{Furthermore, the current coil sensitivity estimation module in SPICER relies on ACS data for calibration of CSM. This dependency limits its utility to Echo Planar Imaging (EPI) sequences, such as fMRI and diffusion MRI. Exploring the expansion of SPICER to incorporate non-ACS data represents an intriguing and valuable direction to enhance its practicality.}

\subsection{Conclusion}
{\bf In conclusion}, we presented SPICER, as a self-supervised MRI reconstruction method that can automatically estimate CSMs to enhance the reconstruction performance even at high undersampled rates.  Our results indicate that the synergistic combination of all three SPICER modules enables it to outperform other self-supervised learning methods and match the performance of the well-known E2E-VarNet trained on fully-sampled groundtruth data.  




\clearpage

\listoffigures


\begin{thebibliography}{10}

\bibitem{griswoldGeneralized2002}
Griswold~MA, Jakob~PM, Heidemann~RM, Nittka~M, Jellus~V, Wang~J, Kiefer~B,
  Haase~A.
\newblock Generalized autocalibrating partially parallel acquisitions
  ({{GRAPPA}}).
\newblock Magn. Reson. Med. 2002; 47:1202--1210.

\bibitem{sodicksonSimultaneous1997}
Sodickson~DK, Manning~WJ.
\newblock Simultaneous acquisition of spatial harmonics ({{SMASH}}): {{Fast}}
  imaging with radiofrequency coil arrays.
\newblock Magn. Reson. Med. 1997; 38:591--603.

\bibitem{pruessmannSENSE1999}
Pruessmann~KP, Weiger~M, Scheidegger~MB, Boesiger~P.
\newblock {{SENSE}}: {{Sensitivity}} encoding for fast {{MRI}}.
\newblock Magn. Reson. Med. 1999; 42:952--962.

\bibitem{ueckerESPIRiTan2014}
Uecker~M, Lai~P, Murphy~MJ, Virtue~P, Elad~M, Pauly~JM, Vasanawala~SS,
  Lustig~M.
\newblock {{ESPIRiT}}-an eigenvalue approach to autocalibrating parallel
  {{MRI}}: {{Where SENSE}} meets {{GRAPPA}}.
\newblock Magn. Reson. Med. 2014; 71:990--1001.

\bibitem{lustigSparse2007}
Lustig~M, Donoho~D, Pauly~JM.
\newblock Sparse {{MRI}}: {{The}} application of compressed sensing for rapid
  {{MR}} imaging.
\newblock Magn. Reson. Med. 2007; 58:1182--1195.

\bibitem{yeCompressed2019}
Ye~JC.
\newblock Compressed sensing {{MRI}}: A review from signal processing
  perspective.
\newblock BMC Biomed. Eng. 2019; 1:8.

\bibitem{ongieDeep2020}
Ongie~G, Jalal~A, Metzler~CA, Baraniuk~RG, Dimakis~AG, Willett~R.
\newblock Deep learning techniques for inverse problems in imaging.
\newblock IEEE J. Sel. Areas Inf. Theory 2020; 1:39--56.

\bibitem{knollDeeplearning2020}
Knoll~F, Hammernik~K, Zhang~C, Moeller~S, Pock~T, Sodickson~DK, Akcakaya~M.
\newblock Deep-learning methods for parallel magnetic resonance imaging
  reconstruction: {{A}} survey of the current approaches, trends, and issues.
\newblock IEEE Signal Process. Mag. 2020; 37:128--140.

\bibitem{montalt2021machine}
MontaltTordera~J, Muthurangu~V, Hauptmann~A, Steeden~J.
\newblock Machine learning in magnetic resonance imaging: image reconstruction.
\newblock Physica Medica 2021; 83:79--87.

\bibitem{Jin.etal2017}
Jin~KH, McCann~MT, Froustey~E, Unser~M.
\newblock Deep {{Convolutional Neural Network}} for {{Inverse Problems}} in
  {{Imaging}}.
\newblock IEEE Trans. on Image Process. 2017; 26:4509--4522.

\bibitem{Zhu.etal2018}
Zhu~B, Liu~JZ, Cauley~SF, Rosen~BR, Rosen~MS.
\newblock Image reconstruction by domain-transform manifold learning.
\newblock Nature 2018; 555:487--492.

\bibitem{hammernikSystematic2021}
Hammernik~K, Schlemper~J, Qin~C, Duan~J, Summers~RM, Rueckert~D.
\newblock Systematic evaluation of iterative deep neural networks for fast
  parallel {{MRI}} reconstruction with sensitivity-weighted coil combination.
\newblock Magn. Reson. Med. 2021; p. mrm.28827.

\bibitem{liangDeep2020}
Liang~D, Cheng~J, Ke~Z, Ying~L.
\newblock Deep magnetic resonance image reconstruction: {{Inverse}} problems
  meet neural networks.
\newblock IEEE Signal Process. Mag. 2020; 37:141--151.

\bibitem{duanVSNet2019}
Duan~J, Schlemper~J, Qin~C, Ouyang~C, Bai~W, Biffi~C, Bello~G, Statton~B,
  O'Regan~DP, Rueckert~D.
\newblock {{VS}}-{{Net}}: {{Variable Splitting Network}} for {{Accelerated
  Parallel MRI Reconstruction}}.
\newblock { In:} Proc. {{Medical Image Computing}} and {{Computer}}-{{Assisted
  Intervention}}, {Cham}, 2019.  pp. 713--722.

\bibitem{aggarwalMoDL2019}
Aggarwal~HK, Mani~MP, Jacob~M.
\newblock {{MoDL}}: {{Model}}-{{Based Deep Learning Architecture}} for
  {{Inverse Problems}}.
\newblock IEEE Trans. Med. Imaging 2019; 38:394--405.

\bibitem{hammernikLearning2018}
Hammernik~K, Klatzer~T, Kobler~E, Recht~MP, Sodickson~DK, Pock~T, Knoll~F.
\newblock Learning a variational network for reconstruction of accelerated
  {{MRI}} data.
\newblock Magn. Reson. Med. 2018; 79:3055--3071.

\bibitem{yaman2020self}
Yaman~B, Hosseini~SAH, Moeller~S, Ellermann~J, U{\u{g}}urbil~K,
  Ak{\c{c}}akaya~M.
\newblock Self-supervised learning of physics-guided reconstruction neural
  networks without fully sampled reference data.
\newblock Magn. Reson. Med. 2020; 84:3172--3191.

\bibitem{arvinteDeep2021}
Arvinte~M, Vishwanath~S, Tewfik~AH, Tamir~JI.
\newblock Deep {{J}}-{{Sense}}: {{Accelerated MRI}} reconstruction via unrolled
  alternating optimization.
\newblock arXiv:2103.02087 [cs, eess] 2021; .

\bibitem{junJoint2021}
Jun~Y, Shin~H, Eo~T, Hwang~D.
\newblock Joint deep model-based {{MR}} image and coil sensitivity
  reconstruction network (joint-icnet) for fast {{MRI}}.
\newblock { In:} Proc. {{IEEE Conf}}. {{Comput}}. {{Vis}}. {{Pattern
  Recognit}}., Nashville, USA, 2021.  pp. 5270--5279.

\bibitem{sriramEndtoend2020}
Sriram~A, Zbontar~J, Murrell~T, Defazio~A, Zitnick~CL, Yakubova~N, Knoll~F,
  Johnson~P.
\newblock End-to-end variational networks for accelerated {{MRI}}
  reconstruction.
\newblock { In:} Proc. {{Medical Image Computing}} and {{Computer}}-{{Assisted
  Intervention}}, Lima, Peru, 2020.  pp. 64--73.

\bibitem{yaman2020selfsupervised}
Yaman~B, Hosseini~SAH, Moeller~S, Ellermann~J, U{\u{g}}urbil~K,
  Ak{\c{c}}akaya~M.
\newblock Self-supervised physics-based deep learning mri reconstruction
  without fully-sampled data.
\newblock { In:} Proc. Int. Symp. Biomedical Imaging, Virtual, 2020.  pp.
  921--925.

\bibitem{akcakayaUnsupervised2021}
Ak{\c c}akaya~M, Yaman~B, Chung~H, Ye~JC.
\newblock Unsupervised {{Deep Learning Methods}} for {{Biological Image
  Reconstruction}}.
\newblock arXiv:2105.08040 [physics] 2021; .

\bibitem{lehtinenNoise2Noise2018}
Lehtinen~J, Munkberg~J, Hasselgren~J, Laine~S, Karras~T, Aittala~M, Aila~T.
\newblock {{Noise2Noise}}: {{Learning}} image restoration without clean data.
\newblock { In:} Proc. {{Int}}. {{Conf}}. {{Mach}}. {{Learn}}., Stockholm,
  Sweden, 2018.

\bibitem{eldenizPhase2Phase2021}
Eldeniz~C, Gan~W, Chen~S, Fraum~TJ, Ludwig~DR, Yan~Y, Liu~J, Vahle~T,
  Krishnamurthy~U, Kamilov~US, An~H.
\newblock {{Phase2Phase}}: {{Respiratory Motion}}-{{Resolved Reconstruction}}
  of {{Free}}-{{Breathing Magnetic Resonance Imaging Using Deep Learning
  Without}} a {{Ground Truth}} for {{Improved Liver Imaging}}.
\newblock Invest. Radiol. 2021; Publish Ahead of Print.

\bibitem{ganDeep2021}
Gan~W, Sun~Y, Eldeniz~C, Liu~J, An~H, Kamilov~US.
\newblock Deep {{Image Reconstruction Using Unregistered Measurements Without
  Groundtruth}}.
\newblock { In:} Proc. {{Int}}. {{Symp}}. {{Biomedical Imaging}}, {Nice,
  France}, April 2021 pp. 1531--1534.

\bibitem{liuRARE2020}
Liu~J, Sun~Y, Eldeniz~C, Gan~W, An~H, Kamilov~US.
\newblock {{RARE}}: {{Image Reconstruction Using Deep Priors Learned Without
  Groundtruth}}.
\newblock IEEE J. Sel. Top. Signal Process. 2020; 14:1088--1099.

\bibitem{gan2021ss}
Gan~W, Hu~Y, Eldeniz~C, Liu~J, Chen~Y, An~H, Kamilov~US.
\newblock Ss-jircs: Self-supervised joint image reconstruction and coil
  sensitivity calibration in parallel mri without ground truth.
\newblock { In:} Proc. IEEE Int. Conf. Comp. Vis. Workshops, Montreal, BC,
  Canada, 2021.  pp. 4048--4056.

\bibitem{block2007undersampled}
Block~KT, Uecker~M, Frahm~J.
\newblock Undersampled radial mri with multiple coils. iterative image
  reconstruction using a total variation constraint.
\newblock Magn. Reson. Med. 2007; 57:1086--1098.

\bibitem{wangDeep2020}
Wang~G, Ye~JC, DeMan~B.
\newblock Deep learning for tomographic image reconstruction.
\newblock Nat. Mach. Intell. 2020; 2:737--748.

\bibitem{ronnebergerUnet2015}
Ronneberger~O, Fischer~P, Brox~T.
\newblock U-net: {{Convolutional}} networks for biomedical image segmentation.
\newblock { In:} Proc. {{Medical Image Computing}} and {{Computer}}-{{Assisted
  Intervention}}, Munich, Germany, 2015.  pp. 234--241.

\bibitem{leeDeep2017}
Lee~D, Yoo~J, Ye~JC.
\newblock Deep residual learning for compressed sensing {{MRI}}.
\newblock { In:} Proc. {{Int}}. {{Symp}}. {{Biomedical Imaging}}, {Melbourne,
  Australia}, April 2017 pp. 15--18.

\bibitem{wangAccelerating2016}
Wang~S, Su~Z, Ying~L, Peng~X, Zhu~S, Liang~F, Feng~D, Liang~D.
\newblock Accelerating magnetic resonance imaging via deep learning.
\newblock { In:} Proc. {{Int}}. {{Symp}}. {{Biomedical Imaging}}, {Prague,
  Czech Republic}, April 2016 pp. 514--517.

\bibitem{zhuImage2018}
Zhu~B, Liu~JZ, Cauley~SF, Rosen~BR, Rosen~MS.
\newblock Image reconstruction by domain-transform manifold learning.
\newblock Nature 2018; 555:487--492.

\bibitem{hanKSpace2020}
Han~Y, Sunwoo~L, Ye~JC.
\newblock K-{{Space Deep Learning}} for {{Accelerated MRI}}.
\newblock IEEE Trans. Med. Imaging 2020; 39:377--386.

\bibitem{Venkatakrishnan.etal2013}
Venkatakrishnan~SV, Bouman~CA, Wohlberg~B.
\newblock Plug-and-{{Play}} priors for model based reconstruction.
\newblock { In:} Proc. {{IEEE Glob}}. {{Conf}}. {{Signal Process}}. {{Inf}}.
  {{Process}}., Austin, USA, December 2013 pp. 945--948.

\bibitem{Kamilov.etal2022}
Kamilov~US, Bouman~CA, Buzzard~GT, Wohlberg~B.
\newblock Plug-and-play methods for integrating physical and learned models in
  computational imaging.
\newblock IEEE Signal Process. Mag. 2022; .
\newblock arXiv:2203.17061.

\bibitem{Romano.etal2017}
Romano~Y, Elad~M, Milanfar~P.
\newblock The {{Little Engine That Could}}: {{Regularization}} by {{Denoising}}
  ({{RED}}).
\newblock SIAM J. Imaging Sci. 2017; 10:1804--1844.

\bibitem{Schlemper.etal2018}
Schlemper~J, Caballero~J, Hajnal~JV, Price~AN, Rueckert~D.
\newblock A {{Deep Cascade}} of {{Convolutional Neural Networks}} for {{Dynamic
  MR Image Reconstruction}}.
\newblock IEEE Trans. Med. Imaging 2018; 37:491--503.

\bibitem{Hammernik.etal2018}
Hammernik~K, Klatzer~T, Kobler~E, Recht~MP, Sodickson~DK, Pock~T, Knoll~F.
\newblock Learning a variational network for reconstruction of accelerated
  {{MRI}} data.
\newblock Magn. Reson. Med. 2018; 79:3055--3071.

\bibitem{Aggarwal.etal2019}
Aggarwal~HK, Mani~MP, Jacob~M.
\newblock {{MoDL}}: {{Model-Based Deep Learning Architecture}} for {{Inverse
  Problems}}.
\newblock IEEE Trans. Med. Imaging 2019; 38:394--405.

\bibitem{liuSGDNet2021}
Liu~J, Sun~Y, Gan~W, Xu~X, Wohlberg~B, Kamilov~US.
\newblock {{SGD}}-{{Net}}: {{Efficient Model}}-{{Based Deep Learning With
  Theoretical Guarantees}}.
\newblock IEEE Trans. Comput. Imaging 2021; 7:598--610.

\bibitem{akccakaya2019scan}
Ak{\c c}akaya~M, Steen~M, Sebastian~W, K{\^a}mil~U.
\newblock Scan-specific robust artificial-neural-networks for k-space
  interpolation (raki) reconstruction: Database-free deep learning for fast
  imaging.
\newblock Magn. Reson. Med. 2019; 81:439--453.

\bibitem{yingJoint2007}
Ying~L, Sheng~J.
\newblock Joint image reconstruction and sensitivity estimation in {{SENSE}}
  ({{JSENSE}}).
\newblock Magn. Reson. Med. 2007; 57:1196--1202.

\bibitem{ueckerImage2008}
Uecker~M, Hohage~T, Block~KT, Frahm~J.
\newblock Image reconstruction by regularized nonlinear inversion-{{Joint}}
  estimation of coil sensitivities and image content.
\newblock Magn. Reson. Med. 2008; 60:674--682.

\bibitem{peng2022deepsense}
Peng~X, Sutton~BP, Lam~F, Liang~Z.
\newblock Deepsense: Learning coil sensitivity functions for sense
  reconstruction using deep learning.
\newblock Magnetic Resonance in Medicine 2022; 87:1894--1902.

\bibitem{yiasemis2022recurrent}
Yiasemis~G, Sonke~J, S{\'a}nchez~C, Teuwen~J.
\newblock Recurrent variational network: A deep learning inverse problem solver
  applied to the task of accelerated mri reconstruction.
\newblock { In:} Proc. {{IEEE Conf}}. {{Comput}}. {{Vis}}. {{Pattern
  Recognit}}., New Orleans, USA, 2022.  pp. 732--741.

\bibitem{blumenthal2022nlinv}
Blumenthal~M, Luo~G, Schilling~M, Haltmeier~M, Uecker~M.
\newblock Nlinv-net: Self-supervised end-2-end learning for reconstructing
  undersampled radial cardiac real-time data.
\newblock { In:} ISMRM annual meeting, London, UK, 2022.

\bibitem{tang2023jsense}
Tang~L, Zhao~Y, Li~Y, Guo~R, Cai~B, Wang~J, Li~Y, Liang~ZP, Peng~X, Luo~J.
\newblock Jsense-pro: Joint sensitivity estimation and image reconstruction in
  parallel imaging using p re-learned subspaces of coil sensitivity functions.
\newblock Magn. Reson. Med. 2023; 89:1531--1542.

\bibitem{meng2019prior}
Meng~N, Yang~Y, Xu~Z, Sun~J.
\newblock A prior learning network for joint image and sensitivity estimation
  in parallel mr imaging.
\newblock { In:} Proc. {{Medical Image Computing}} and {{Computer}}-{{Assisted
  Intervention}}, Shenzhen, China, 2019.  pp. 732--740.

\bibitem{zeng2021review}
Zeng~G, Guo~Y, Zhan~J, Wang~Z, Lai~Z, Du~X, Qu~X, Guo~D.
\newblock A review on deep learning mri reconstruction without fully sampled
  k-space.
\newblock BMC Medical Imaging 2021; 21:1--11.

\bibitem{Akcakaya.etal2021}
Ak{\c{c}}akaya~M, Yaman~B, Chung~H, Ye~JC.
\newblock Unsupervised deep learning methods for biological image
  reconstruction and enhancement: An overview from a signal processing
  perspective.
\newblock IEEE Signal Process. Mag. 2022; 39:28--44.

\bibitem{Tachella.etal2022a}
Tachella~J, Chen~D, Davies~M.
\newblock Sampling {{Theorems}} for {{Unsupervised Learning}} in {{Linear
  Inverse Problems}}.
\newblock arXiv:2203.12513 2022; .

\bibitem{Lehtinen.etal2018}
Lehtinen~J, Munkberg~J, Hasselgren~J, Laine~S, Karras~T, Aittala~M, Aila~T.
\newblock {{Noise2Noise}}: {{Learning}} image restoration without clean data.
\newblock { In:} Proc. {{Int}}. {{Conf}}. {{Mach}}. {{Learn}}., Stockholm,
  Sweden, 2018.

\bibitem{Krull.etal2019}
Krull~A, Buchholz~TO, Jug~F.
\newblock {{Noise2Void}} - {{Learning Denoising From Single Noisy Images}}.
\newblock { In:} Proc. {{IEEE Conf}}. {{Comput}}. {{Vis}}. {{Pattern
  Recognit}}., Long Beach, USA, June 2019 pp. 2124--2132.

\bibitem{Ulyanov.etal2018}
Ulyanov~D, Vedaldi~A, Lempitsky~V.
\newblock Deep image prior.
\newblock { In:} Proc. {{IEEE Conf}}. {{Comput}}. {{Vis}}. {{Pattern
  Recognit}}., Salt Lake City, USA, 2018.  pp. 9446--9454.

\bibitem{bora2018ambientgan}
Bora~A, Price~E, Dimakis~AG.
\newblock Ambientgan: Generative models from lossy measurements.
\newblock { In:} Int. Conf. Learn. Represent., Vancouver, Canada, 2018.

\bibitem{chen2021equivariant}
Chen~D, Tachella~J, Davies~ME.
\newblock Equivariant imaging: Learning beyond the range space.
\newblock { In:} Proc. IEEE Conf. Comput. Vis. Pattern Recognit., Nashville,
  USA, 2021.  pp. 4379--4388.

\bibitem{huang2022unsupervised}
Huang~P, Zhang~C, Zhang~X, Li~X, Dong~L, Ying~L.
\newblock Unsupervised deep unrolled reconstruction using regularization by
  denoising.
\newblock arXiv preprint arXiv:2205.03519 2022; .

\bibitem{NEURIPS2019_bdbca288}
Paszke~A, Gross~S, Massa~F, Lerer~A, Bradbury~J, Chanan~G, Killeen~T, Lin~Z,
  Gimelshein~N, Antiga~L, Desmaison~A, Kopf~A, Yang~E, DeVito~Z, Raison~M,
  Tejani~A, Chilamkurthy~S, Steiner~B, Fang~L, Bai~J, Chintala~S.
\newblock Pytorch: An imperative style, high-performance deep learning library.
\newblock { In:} Proc. Ann. Conf. Neural Information Processing Systems,
  Vancouver, Canada, 2019.

\bibitem{benjamini1995controlling}
Benjamini~Y, Hochberg~Y.
\newblock Controlling the false discovery rate: a practical and powerful
  approach to multiple testing.
\newblock Journal of the Royal statistical society: series B (Methodological)
  1995; 57:289--300.

\end{thebibliography}
\end{document}